\documentclass[12pt]{report}
\usepackage[cp866]{inputenc}
\usepackage[english]{babel}

\usepackage{graphicx}
\usepackage{amsfonts}
\usepackage{amssymb}
\usepackage{amsmath}
\usepackage[colorlinks,urlcolor=blue]{hyperref}

\begin{document}

\title{\bf Hamiltonian dynamics of a symmetric top in external fields having axial symmetry. Levitating Orbitron}
\providecommand{\keywords}[1]{\textbf{\textit{Index terms---}} #1}

\author{
  Stanislav S. Zub\\
  Faculty of Cybernetics,\\
  Taras Shevchenko National University of Kyiv,\\
  Glushkov boul., 2, corps 6.,\\
  Kyiv, Ukraine 03680\\
  \texttt{stah@univ.kiev.ua}\\
  \\
  Sergiy I. Zub\\
  Institute of Metrology,\\
  Mironositskaya st., 42,\\
  Kharkiv, Ukraine 61002\\
  \texttt{sergii.zub@gmail.com}}

\maketitle

\begin{abstract}
The symmetric top is a special case of the general top,
and canonical Poisson structure on $T^*SE(3)$ is
the common method of its description.
This structure is invariant under the right action of $SO(3)$,
but the Hamiltonian of the symmetric top is invariant
only under the right action of subgroup $S^1$
that corresponds to the rotation
around the symmetry axis of the symmetric top.
So, its Poisson structure was obtained as the reduction $T^*SE(3)/S^1$.
Next we propose the Hamiltonian that describes the wide class of the interaction models
of symmetric top and axially-symmetric external field.
The stability of the levitating Orbitron in relative equilibrium was proved.

.
\end{abstract}

\tableofcontents

%
\inputencoding{cp1251}
\newcommand{\s}[1]{\ensuremath{\boldsymbol{\sigma}_{#1}}}
\newcommand{\bsym}[1]{\ensuremath{\boldsymbol{#1}}}
\newcommand{\w}{\ensuremath{\boldsymbol{\wedge}}}
\newcommand{\lc}{\ensuremath{\boldsymbol{\rfloor}}}
\newcommand{\rc}{\ensuremath{\boldsymbol{\lfloor}}}

\thispagestyle{empty}

\newpage
\chapter{Reduction of $T^*SE(3)$ to the Poisson structure for a symmetric top}

\section{Introduction}
\label{intro}

\bigskip
The model of the Lagrange top has a long history and it is still the subject of the investigations \cite{LeRaSiMa92}.
The widespread use of the group-theoretic methods of Hamiltonian mechanics gave the new impulse
for the study of this classical model \cite{LeRaSiMa92, MarRat98}.
The effectiveness of this model emerged in the study of the stability of the magnetic dynamical systems \cite{ZubOrb13,DullinChDyn04}.
As it will be seen below the Hamiltonian reduction from the general asymmetric body to the symmetric top leads
to the Lie-Poisson structure, embedded in $\bsym{se(3)}^*$.
The symplectic leaves of this structure are the orbits of the coadjoint representation of $SE(3)$ group.

\section{ Hamiltonian formalism on $T^*SO(3)$}
\label{Symplectic}

\bigskip
\subsection{ Representation of the right trivialization for $T^*SO(3)$}

Let's look at some useful relations for group $SO(3)$ and its cotangent bundle,
many of which can be found in \cite{AbrMar02,MarRat98}.

Group $SO(3)$ is formed by orthogonal, unimodular matrices $\mathbf{R}$ i.e.
$\mathbf{R}^T = \mathbf{R}^{-1}$, $\rm det(\mathbf{R})= 1$.
Correspondingly the Lie algebra $\bsym{so(3)}$ is formed by $3\times 3$-antisymmetric matrices
with Lie bracket in the form of matrix commutator.
Let's consider a vector space isomorphism $\hat{}:\mathbb{R}^3\rightarrow\bsym{so(3)}$
is such that \cite[p. 285]{MarRat98}:
$\widehat{\bsym{\xi}}_{kl} = -\varepsilon_{ikl}\xi_i$, $\xi_i = -\frac12 \varepsilon_{ikl} \widehat{\bsym{\xi}}_{kl}$,
where $\varepsilon_{ikl}$ --- Levi-Civita symbol. Then
\[
\begin{cases}
   \widehat{\bsym{\xi}}\bsym{\eta} = \bsym{\xi}\times \bsym{\eta};\\
   [\widehat{\bsym{\xi}},\widehat{\bsym{\eta}}] =
   \widehat{\bsym{\xi}}\widehat{\bsym{\eta}} - \widehat{\bsym{\eta}}\widehat{\bsym{\xi}} =
   \widehat{\bsym{\xi}\times \bsym{\eta}};\\
   \langle \bsym{\xi},\bsym{\eta} \rangle
   = -\frac12 {\rm tr}(\widehat{\bsym{\xi}}\widehat{\bsym{\eta}});\\
   \mathbf{B}\widehat{\bsym{\xi}}\mathbf{B}^{-1} = \widehat{\mathbf{B}\bsym{\xi}}
\end{cases}
\label{A1}
\eqno(1.2.1.1)\]
The scalar product introduced above allows us to state the equivalence
of the Lie algebra and its dual space $\bsym{so(3)}^*\simeq\bsym{so(3)}$.
Symbol ``$\simeq$'' defines diffeomorphism.
As a rule, the diffeomorphisms used below have a simple group-theoretical
or differential-geometric sense that is explained in the cited literature.

In the representation of the {\it right trivialization} which is shown in \cite{LeRaSiMa92}
and \cite[p. 314]{AbrMar02} corresponds to the {\it inertial system}, we have:\\
$T SO(3)\simeq SO(3)\times\bsym{so(3)}$, $T^*SO(3)\simeq SO(3)\times\bsym{so(3)}^*$.

Then the right and the left actions of group $SO(3)$ on $T^*SO(3)$
(Cotangent Lift \cite[p. 166]{MarRat98}) have the form
\[
\begin{cases}
      R^{ct}_{\mathbf{B}}: (\mathbf{R},\widehat{\bsym{\pi}})\in T^*SO(3)
      \rightarrow(\mathbf{R}\mathbf{B},\widehat{\bsym{\pi}}), \quad \mathbf{B}\in SO(3);\\
      L^{ct}_{\mathbf{B}}: (\mathbf{R},\widehat{\bsym{\pi}})\in T^*SO(3)
      \rightarrow(\mathbf{B}\mathbf{R},\mathbf{B}\widehat{\bsym{\pi}}\mathbf{B}^{-1})
      = (\mathbf{B}\mathbf{R},{\rm Ad}^*_{\mathbf{B}^{-1}}\widehat{\bsym{\pi}})
\end{cases}
\label{A2}
\eqno(1.2.1.2)\]

\subsection{ Symplectic and Poisson structures on $T^*SO(3)$ \cite{AbrMar02,LeRaSiMa92,ZubLie13,ZubRBDDAN13}}

Liouville form on $T^*SO(3)\simeq SO(3)\times\bsym{so(3)}^*$ looks like
\[
   \Theta^{T^*SO(3)}_{|(\mathbf{R},\bsym{\pi})}
   = -\frac12 {\rm tr}(\widehat{\bsym{\pi}}\widehat{\bsym{\delta R}})
   = \pi_i\delta R^i,
\label{A3}
\eqno(1.2.2.1)\]
where $\widehat{\bsym{\delta R}} = (d\mathbf{R})\mathbf{R}^{-1}$ is the right-invariant of Maurer-Cartan 1-form.

Then by using the Maurer-Cartan equation \cite[p. 276]{MarRat98} for the canonical symplectic 2-form we have
\[
   \Omega^{T^*SO(3)}_{can}  = -d\Theta^{T^*SO(3)} = {\delta R}^i\w d\pi_i - \pi_i [\delta R,\delta R]^i =
\label{A4}
\eqno(1.2.2.2)\]
\[ = -\frac12\varepsilon_{ijk}\delta R_{jk}\w d \pi_i
   + \frac12 \pi_i\varepsilon_{ijk}\delta R_{js}\w \delta R_{sk}
\]

Any given symplectic structure $\Omega$ defines a Poisson structure on the same manifold as follows
\[
   \{F, G\}(z) = \Omega(\xi_F(z),\xi_G(z))
   = \partial_{\xi_G}F = -\partial_{\xi_F}G,
\label{A5}
\eqno(1.2.2.3)\]
where equation $i_{\xi_G}\Omega = dG$ is performed for vector field $\xi_G$.

By considering the elements of the matrix $\mathbf{R}$ and the components of momentum $\bsym{\pi}$
as the dynamic variables on $T^*SO(3)$, we can obtain the following set of Poisson brackets
that completely defines a Poisson structure on $T^*SO(3)$:
\[
   \{R_{ij}, R_{kl}\} = 0, \quad
   \{\pi_i, R_{jk}\} = \varepsilon_{ijl}R_{lk}, \quad
   \{\pi_i, \pi_j\} = \varepsilon_{ijl}\pi_l.
\label{A6}
\eqno(1.2.2.4)\]

Note that in the inertial system Poisson brackets for the matrix elements $\mathbf {R}$
are grouped in columns. For example, Poisson brackets for the elements of the 3d-column is expressed
through the elements of the 3d-column only.

\section{ Reduction of the Poisson structure for a symmetric top}
\label{S1Reduction}

Poisson structure (1.2.2.4) is invariant under right translations of constant matrix $\mathbf{B} \in SO(3)$:
\[
   \{(R B)_{ij}, (R B)_{kl}\} = 0, \quad
   \{\pi_i, (R B)_{jn}\} = \varepsilon_{ijl}(R B)_{ln}, \quad
   \{\pi_i, \pi_j\} = \varepsilon_{ijl}\pi_l.
\label{B1}
\eqno(1.3.1)\]

Subgroup $S^1~\in~SO(3)$ is of interest for the symmetric top
where $S^1 = \{ \mathbf{Z}\in SO(3): Z_{i3}= \delta_{i3} \}$
(it is assumed that the axis of the body symmetry is directed along the vector $\bsym{E}_3$)
in the body frame).
Then $(\mathbf{R}\mathbf{Z})_{i3} = R_{ij}Z_{j3} = R_{i3}$,
i.e. 3d-row of the matrix $\mathbf{R}$ remains invariant under right translations corresponding to subgroup $S^1$.

Let's consider projection $SO(3)$ on the sphere $S^2$ (as a set of unit vectors $\bsym{\nu}$ with $\bsym{\nu}^2=1$)
\[
   \tau: \mathbf{R}\mapsto \bsym{\nu} = R_{i3}\bsym{e}_i, \quad \nu_i = R_{i3},
\label{B2}
\eqno(1.3.2)\]
where $\bsym{e}_i$ are basis vectors for inertial system.

This projection generates a map
\[
   \tilde{\tau}: T^*SO(3)\ni (\mathbf{R},\bsym{\pi})
   \mapsto (\tau(\mathbf{R}),\bsym{\pi}) = (\bsym{\nu},\bsym{\pi})\in K_1,
\label{B3}
\eqno(1.3.3)\]
where $K_1\simeq S^2\times \bsym{so(3)}^*\subset \mathbb{R}^3\times \bsym{so(3)}^*\simeq\bsym{se(3)}^* = (\mathbb{R}^3\circledS\bsym{so(3)}^*$).

On $\bsym{se(3)}^*$, as on any space that is dual to the Lie algebra,
there exists a canonical Lie-Poisson structure \cite[p. 425]{MarRat98}.
In this case \cite[pp. 491,367]{MarRat98} this structure is determined by the following Poisson brackets
\[
   \{\nu_i, \nu_k\} = 0, \quad
   \{\pi_i, \nu_j\} = \varepsilon_{ijl}\nu_l, \quad
   \{\pi_i, \pi_j\} = \varepsilon_{ijl}\pi_l.
\label{B4}
\eqno(1.3.4)\]

Comparing (1.3.4) with (1.2.2.4) shows that a surjective mapping $\tilde{\tau}$ is poissonian.
Thus the conditions of {\it Theorem 10.5.1} \cite[p. 355]{MarRat98} are satisfied, and hence,
\[
   K_1 \simeq T^*SO(3)/ S^1
\label{B5}
\eqno(1.3.5)\]

If Hamiltonian $H$ on $T^*SO(3)$ is $S^1$ is invariant,
then there is a Hamiltonian $h$ on $K_1$ that exists in such a way that $H=h\circ \tilde{\tau}$
and with trajectories of the dynamic system with Hamiltonian $H$ are
$\tilde{\tau}$-associated with trajectories for Hamiltonian $h$ \cite[p. 355]{MarRat98}.

\section{ The structure of symplectic leaves for the dynamics of a symmetric top}
\label{Leaves}

{\bf Proposition 1.} {\it Let's consider the functions on $\bsym{se(3)}^*$
\[
   C_1(\bsym{\nu},\bsym{\pi}) = \bsym{\nu}^2, \quad
   C_2(\bsym{\nu},\bsym{\pi}) = \bsym{\nu}\cdot\bsym{\pi}.
\label{C1}
\eqno(1.4.1)\]

1. Functions $C_1,C_2$ are independent in all points $\bsym{se(3)}^*$ for each $\bsym{\nu}\neq 0$.

2. This means that a common level $\textsl{L}_{(\bsym{\nu}_0,\bsym{\pi}_0)}$ for functions (1.4.1) when
$\bsym{\nu}_0\neq 0$ is a submanifold in $\bsym{se(3)}^*$.

3. There is such coordinate system $(c^1,c^2,u^1,\dots,u^4)$ in the neighborhood
of an any point $(\bsym{\nu}_0,\bsym{\pi}_0)\in \bsym{se(3)}^* (\bsym{\nu}_0^2>0)$ that
in this neighborhood $\textsl{L}_{(\bsym{\nu},\bsym{\pi})}$ is determined by equations
\[
\begin{cases}
   c^1 = C_1(\bsym{\nu},\bsym{\pi});\\
   c^2 = C_2(\bsym{\nu},\bsym{\pi});\\
\end{cases}
\label{C2}
\eqno(1.4.2)\]

{\bf Remark 1}. Coordinates $(u^1,\dots,u^4)$ are internal on the submanifold determined
by the fixed values $(c^1,c^2)$.
}

$\square$

1. Functions are independent in the neighborhood of point $(\bsym{\nu},\bsym{\pi})$
if differentials of these functions are independent at this point.

We have
\[
\begin{cases}
      {\rm d}C_1 = \frac{\partial C_1}{\partial_{\nu_k}}{\rm d}\nu_k
                 + \frac{\partial C_1}{\partial_{\pi_k}}{\rm d}\pi_k
                 = 2\nu_k{\rm d}\nu_k;\\
      {\rm d}C_2 = \frac{\partial C_2}{\partial_{\nu_k}}{\rm d}\nu_k
                 + \frac{\partial C_2}{\partial_{\pi_k}}{\rm d}\pi_k
                 = \pi_k{\rm d}\nu_k + \nu_k{\rm d}\pi_k;\\
\end{cases}
\]

Let's assume the contrary, there are such $\alpha,\beta$ that $\alpha {\rm d}C_1 + \beta {\rm d}C_2~=~0$,
then when equating to zero the coefficients of the  differentials $d\pi_k$, we get
\[
 (\bsym{\nu}^2 > 0)\&(\beta\nu_k {\rm d}\pi_k = 0)
  \longrightarrow \beta\nu_k = 0,\forall k
  \longrightarrow \beta = 0
\]

Correspondingly, equating coefficients to zero when the differentials are $d\nu^k$, we get
\[
 ( \bsym{\nu}^2 > 0)\&(\alpha {\rm d}C_1) = 0\longrightarrow
   \alpha \nu_k{\rm d}\nu_k = 0\longrightarrow
   \alpha\nu_k = 0,\forall k
   \longrightarrow\alpha = 0
\]

2. As it was proved in item 1 the map $C=C_1\times C_2$ is regular in points
\noindent
$(\bsym{\nu},\bsym{\pi})\in\bsym{se(3)}^*$, where $\bsym{\nu}\neq 0$,
that is why item 2 is a direct corollary fact of {\bf Submersion Theorem} \cite[p. 175]{MarTA}.

3. This statement is the simple consequence of the {\bf Local Onto Theorem} \cite[p. 175]{MarTA} for mapping $C - C_0$.

$\blacksquare$

\bigskip
Coadjoint action of group $SE(3)$ on $\bsym{se(3)}^*$ looks like ((14.7.10), \cite[p. 492]{MarRat98})
\[
   Ad^*_{(\bsym{a},\mathbf{A})^{-1}}(\bsym{\nu},\bsym{\pi})
   = (\mathbf{A}[\bsym{\nu}],\bsym{a}\times \mathbf{A}[\bsym{\nu}] + \mathbf{A}[\bsym{\pi}]).
\label{C3}
\eqno(1.4.3)\]

{\bf Proposition 2}. {\it Let $K=\bsym{se(3)}^*\backslash$ $\{(\bsym{\nu},\bsym{\pi}):\bsym{\nu}= 0 \}$
that is $K$ is an open submanifold $\bsym{se(3)}^*$ with Lie-Poisson structure (1.3.4)
and coadjoint action (1.4.3) induced from $\bsym{se(3)}^*$, then

1. Functions $C_1,C_2$ are the Casimir functions on $K$.

2. $\textsl{L}_{(\bsym{\nu}_0,\bsym{\pi}_0)}$ is orbit $\textsl{O}_{(\bsym{\nu}_0,\bsym{\pi}_0)}$
of the coadjoint group presentation $SE(3)$.

3. $\textsl{L}_{(\bsym{\nu}_0,\bsym{\pi}_0)}$ is the symplectic leaf $K$,
and every symplectic leaf $K$ can be represented as $\textsl{L}_{(\bsym{\nu}_0,\bsym{\pi}_0)}$.

4. $\textsl{L}_{(\bsym{\nu}_0,\bsym{\pi}_0)}$ is a Poisson submanifold in $K$
}

$\square$

1. Functions (1.4.1) are invariant relative to the coadjoint action (1.4.3),
so on the basis of proposition 12.6.1 (see \cite[p. 421]{MarRat98}) we can assume that $C_1,C_2$
are the Casimir functions.

2. For any point $(\bsym{\nu},\bsym{\pi})\in \textsl{L}_{(\bsym{\nu}_0,\bsym{\pi}_0)}$
there are such parameters $(\bsym{a},\mathbf{A})$ that
$(\bsym{\nu},\bsym{\pi}) = Ad^*_{(\bsym{a},\mathbf{A})^{-1}}(\bsym{\nu}_0,\bsym{\pi}_0)$.
Indeed, $\bsym{\nu}\in S^2_{\nu_0}$ and there is a rotation $\mathbf{A}$ such that
$\bsym{\nu}=\mathbf{A}[\bsym{\nu}_0]$ and a suitable vector $\bsym{a}$ can be found with the formula
$\bsym{a} = \frac1{\nu_0^2}\bsym{\nu}\times\left(\bsym{\pi} - \mathbf{A}[\bsym{\pi}_0]\right)$.

3. As it follows from {\bf Corollary 14.4.3} and {\bf Definition 10.4.3} \cite[p. 477, (i)]{MarRat98}
the connected component of the orbit of the coadjoint presentation is a symplectic leaf.
The orbits of group $SE(3)$ are connected.

4. According to {\bf Proposition 1.2} $\textsl{L}_{(\bsym{\nu}_0,\bsym{\pi}_0)}$
is not only {\it immersed submanifold}, but also {\it Poisson submanifold} (\cite[p. 347]{MarRat98})
$K$ (and $\bsym{se(3)}^*$).

$\blacksquare$

In sections 14.7 \cite{MarRat98} and in section 4.4 \cite{MarMisOrtPerRat07}
the orbits of the coadjoint presentation of group $SE(3)$ were investigated.
Their structure as the symplectic manifolds with 2-form of Kirillov-Kostant-Souriau (KKS)
$\Omega^{\textsl{O}_{(\bsym{\nu}_0,\bsym{\pi}_0)}}_{KKS}$ is described in {\bf Theorem 4.4.1}
(see \cite[p. 142]{MarMisOrtPerRat07}).

We are interested in orbits $\textsl{O}_{(\bsym{\nu}_0,\bsym{\pi}_0)}$
with $\bsym{\nu}_0\neq 0$. As a consequence of this theorem, we obtain for these orbits:

{\bf Proposition 3}. {\it Orbit $\textsl{O}_{(\bsym{\nu}_0,\bsym{\pi}_0)}$ with $\bsym{\nu}_0\neq 0$
is diffeomorphic to the cotangent bundle of the sphere $T^*S_{|\bsym{\nu}_0|}^2$,
and the symplectic form on the orbit differs from the canonical symplectic form
on the cotangent bundle of the sphere in so-called magnetic term.
\[
\begin{cases}
   \textsl{O}_{(\bsym{\nu}_0,\bsym{\pi}_0)} \simeq T^*S_{|\bsym{\nu}_0|}^2;\\
   \Omega^{\textsl{O}_{(\bsym{\nu}_0,\bsym{\pi}_0)}}_{KKS}
   = \Omega^{T^*S_{|\bsym{\nu}_0|}^2}_{can}-\rho^*\textsl{B};\\
   \textsl{B}(\bsym{\xi}\times \bsym{\nu}, \bsym{\eta}\times \bsym{\nu})_{|\bsym{\nu}}
   = -\frac{C_2(\bsym{\nu}_0,\bsym{\pi}_0)}{C_1(\bsym{\nu}_0,\bsym{\pi}_0)}
   \langle \bsym{\xi}\times \bsym{\eta},\bsym{\nu}\rangle,
\end{cases}
\]
where $\rho:T^*S_{|\bsym{\nu}_0|}^2\rightarrow S_{|\bsym{\nu}_0|}^2$ is the projection onto cotangent bundle,

$\textsl{B}$ --- 2-form on sphere,
$\bsym{\nu}\in S_{|\bsym{\nu}_0|}^2$, $\bsym{\xi},\bsym{\eta}\in\mathbb{R}^3$.
}

\section{ Hamiltonian formalism on $T^*SE(3)$}
\label{Hamilton}

\bigskip

\subsection{ Bundle of orthonormal oriented triads $O^{+}(E^3)$ as the configuration space for rigid body dynamics }

Let's consider a frame of reference, associated with the body that is equivalent to an orthonormal frame (triad)
with the origin $x\in E^3$ in the center of mass of a rigid body and unit vectors $\vec{E}_i$,
directed along the principal axes of inertia tensor with the same orientation as the inertial system
$\{\vec{e}_i\}$, ($\vec{e}_1\times\vec{e}_2=\vec{e}_3$).

That means that the configuration space for a rigid body coincides with the bundle
$(O^{+}(E^3),\varrho,E^3)$, $O^{+}(E^3)\xrightarrow{\varrho}E^3$
of the oriented orthonormal frames in the Euclidean space $E^3$.

The elements of $O^{+}(E^3)$ are $z=(x,\{\vec{E}_i\})$,
($\vec{E}_1\times\vec{E}_2=\vec{E}_3$), then $\varrho(z)=x\in E^3$.

On $O^{+}(E^3)$, as on the principal bundle with structure group $SO(3)$
a canonical {\it right} action of group $SO(3)$ is determined:
\[
   r_{\mathbf{B}}: z = (x,\{\vec{E}_i \})\in O^{+}(E^3)\mapsto (x,\{ B^k{}_i \vec{E}_k \}),
   \quad \mathbf{B}\in SO(3)
\label{D1}
\eqno(1.5.1.1)\]

In addition, there is {\it left} action of $SE(3)$ for {\it flat Euclidean space} $E^3$.
It is not canonical in terms of the general theory of the fiber bundles. 

Let's choose a fixed point $O \in E^3$ and a fixed triad,
forming a Cartesian system of reference $z_0 = (O,\{\vec{e}_i\})$,
then each point $x\in E^3$ is presented by a radius-vector $\bsym{x}$,
and each rotation $\mathbf{A}\in SO(3)$ is presented by matrix $A_{ki}$
such that $\mathbf{A}\vec{e}_i=A_{ki}\vec{e}_k$.

Then the {\it left} action of $SE(3)$ for $z\in O^{+}(E^3)$ looks like
\[
   l_{(\bsym{a},\mathbf{A})}z
   = (\bsym{a} + \mathbf{A}\bsym{x},\{\mathbf{A}\vec{E}_i\})
\label{D2}
\eqno(1.5.1.2)\]

This action $SE(3)$ is simply transitive, i.e. any element
$z\in O^{+}(E^3)$ can be obtained by the {\it left} action (1.5.1.2)
from $z_0\in~O^{+}(E^3)$ by the only way
\[
   z = l_{(\bsym{x},\mathbf{R})}z_0,
   \quad R_{ik} = \langle\vec{E}_k,\vec{e}_i\rangle
\label{D3}
\eqno(1.5.1.3)\]

Mapping
\[
   \Psi: z =(x,\{\vec{E}_i\}) \mapsto (\bsym{x},\mathbf{R}),
   \quad R_{ik} = \langle \vec{E}_k, \vec{e}_i\rangle
\label{D4}
\eqno(1.5.1.4)\]
is a global map of the bundle $O^{+}(E^3)$, actions (1.5.1.1) and (1.5.1.2)
in this map are the right and {\it left} translations on $SE(3)$ correspondingly.
\[
   \Psi\circ r_{\mathbf{B}}\circ \Psi^{-1} = R_{\mathbf{B}},\quad
   \Psi\circ l_{(\bsym{a},\mathbf{A})}\circ \Psi^{-1} = L_{(\bsym{a},\mathbf{A})}
\label{D5f}
\eqno(1.5.1.5)\] 

Diffeomorphism $\Psi$ allowes to consider group $SE(3)$ as a configuration space
for the rigid body dynamics that will be assumed further.

\bigskip
\subsection{ Poisson and symplectic structures on $T^*SE(3)$ in the inertial system}

Take into consideration that group $SE(3)$ as a manifold is a direct product, we have
\[
   T^*(SE(3)) \simeq T^*(\mathbb{R}^3\times SO(3))
   \simeq T^*(\mathbb{R}^3)\times T^*(SO(3))
\label{E1}
\eqno(1.5.2.1)\]

Representations (1.5.2.1) are a direct product \cite[p. 81-82]{Souriau97},
where translational and rotational degrees of freedom are separated from each other
in the symplectic and Poisson structures.
\[
\Omega^{T^*SE(3)}_{can} = \Omega^{T^*R^3}_{can} + \Omega^{T^*SO(3)}_{can}
\label{E2}
\eqno(1.5.2.2)\]
\[ = d x^i\w d p_i -\frac12\varepsilon_{ijk}\delta R_{jk}\w d \pi_i
   + \frac12 \pi_i\varepsilon_{ijk}\delta R_{js}\w \delta R_{sk}
\]

Poisson brackets that correspond $\Omega^{T^*SE(3)}_{can}$ have the form:
\[
\{x_i, p_j\} = \delta_{ij}, \quad
   \{\pi_i, \pi_j\} = \varepsilon_{ijl}\pi_l, \quad
   \{\pi_i, R_{jk}\} = \varepsilon_{ijl}R_{lk}.
\label{E3}
\eqno(1.5.2.3)\]

Poisson brackets (1.5.2.3) explain that the basic dynamic variables refer to the {\it inertial system}.

Let $R^{ct}_{\mathbf{B}}$ --- Cotangent Lift of the {\it right} translation $R_{\mathbf{B}}$ on group $SE(3)$
by elements $(0,\mathbf{B})\in SO(3)$, $L^{ct}_{(\bsym{b},\mathbf{B})}$ --- Cotangent Lift
of the {\it left} translation respectively.

In the inertial system we have, expanding (1.2.1.2)
\[
\begin{cases}
      (\bsym{b},\mathbf{B})\in SE(3),\quad
      ((\bsym{x},\bsym{p}),(\mathbf{R},\bsym{\pi}))\in T^*SE(3);\\
      R_{\mathbf{B}}: (\bsym{x},\mathbf{R})
      \rightarrow(\bsym{x},\mathbf{R}\mathbf{B});\\
      L_{(\bsym{a},\mathbf{A})}: (\bsym{x},\mathbf{R})
      \rightarrow(\bsym{a}+\mathbf{A}\bsym{x},\mathbf{A}\mathbf{R});\\
      R^{ct}_{\mathbf{B}}: ((\bsym{x},\bsym{p}),(\mathbf{R},\bsym{\pi}))
      \rightarrow((\bsym{x},\bsym{p}),(\mathbf{R}\mathbf{B},\bsym{\pi}));\\
      L^{ct}_{(\bsym{a},\mathbf{A})}: ((\bsym{x},\bsym{p}),(\mathbf{R},\bsym{\pi}))
      \rightarrow((\bsym{a}+\mathbf{A}\bsym{x},\mathbf{A}\bsym{p}),
      (\mathbf{A}\mathbf{R}, \mathbf{A}\bsym{\pi}))\\
\end{cases}
\label{E4}
\eqno(1.5.2.4)\]

In order to make a transition for the reference frame that is connected with the body,
it is necessary, in accordance with the last line in (1.5.2.4)
to make the following canonical transformation of the dynamic variables.
\[
\begin{cases}
   \bsym{P} =  \mathbf{R}^{-1}\bsym{p}; \\
   \bsym{\Pi} = \mathbf{R}^{-1}\bsym{\pi}; \\
   \bsym{X} = \bsym{x},\quad \mathbf{R} = \mathbf{R}
\end{cases}
\label{E5}
\eqno(1.5.2.5)\]
where $\bsym{p}$ is a body momentum in the inertial system;

$\bsym{\pi}$ is a self angular momentum of the body in the inertial system;

$\bsym{P}$ is a body momentum in the body frame;

$\bsym{\Pi}$ is a self angular momentum of the body in the body frame.

If we insert expression (1.5.2.5) into (1.5.2.3),
we get the following Poisson brackets (only nonzero) in the body frame
\[
\begin{cases}
   \{X_i, P_j\} = R_{ij}, \qquad
   \{\Pi_i, P_j\} = -\varepsilon_{ijk}P_k, \\
   \{\Pi_i, \Pi_j\} = -\varepsilon_{ijk}\Pi_k,
   \{\Pi_k, R_{ij} \} = -\varepsilon_{kjl}R_{il}.
\end{cases}
\label{E6}
\eqno(1.5.2.6)\]

{\bf Remark 2}. We will write the {\it arithmetic} vectors that represent
the physical vector components in the inertial system in small bold letters
and the components of the same vector in the body frame in capital bold letters.
For example,
\[
   \bsym{\pi} =
   \begin{bmatrix}
      \langle \vec{e}_1, \vec{\pi} \rangle \\
      \langle \vec{e}_2, \vec{\pi} \rangle \\
      \langle \vec{e}_3, \vec{\pi} \rangle)
   \end{bmatrix},\qquad \bsym{\Pi} =
   \begin{bmatrix}
      \langle \vec{E}_1, \vec{\pi} \rangle \\
      \langle \vec{E}_2, \vec{\pi} \rangle \\
      \langle \vec{E}_3, \vec{\pi} \rangle)
   \end{bmatrix}
\]

{\bf Remark 3}. Expressions (1.5.2.2,1.5.2.3) for the symplectic and Poisson structures
do not correspond to the {\it right} trivialization $T^*SE(3)$, described in \cite{GOZ},
the {\it right} trivialization in these expressions affects only the rotational degrees of freedom
from $T^*SO(3)$, and (1.5.2.5) fully corresponds to the {\it left} trivialization $T^*SE(3)$ \cite{GOZ}.

This is coordinated with understanding of $O^{+}(E^3)$
as a configuration space for rigid body dynamics
and the roles of the right and left actions (1.5.1.1,1.5.1.2).

{\bf Remark 4}. The map of bundle $O^{+}(E^3)$ corresponding to the inertial system
has the advantage that translational and rotational degrees of freedom are divided
and it is fully realized in the model of a symmetric top.
As for general top, the kinetic rotation energy is the left-invariant function on $T^*SE(3)$
and therefore body frame is traditionally used.
For general top we can combine the advantages of these two maps,
if we use transformation (1.5.2.7) instead of (1.5.2.5)
\[
\begin{cases}
   \bsym{\Pi} = \mathbf{R}^{-1}\bsym{\pi}; \\
   \bsym{P} = \bsym{p},\quad \bsym{X} = \bsym{x},\quad \mathbf{R} = \mathbf{R}
\end{cases}
\label{E7}
\eqno(1.5.2.7)\]
Using of this map is described in \cite{ZubRBDDAN13}.

\section{ Reduction of $T^*SE(3)$ to the Poisson structure for a symmetric top}
\label{Leaves}

As in case of $T^*SO(3)$, the Poisson brackets for $T^*SE(3)$ are invariant for the {\it right} translations
\[
   R_{\mathbf{B}}:((\bsym{x},\bsym{p}),(\mathbf{R},\bsym{\pi}))
   \mapsto ((\bsym{x},\bsym{p}),(\mathbf{R}\mathbf{B},\bsym{\pi})),
   \quad \mathbf{B} \in SO(3).
\label{D5}
\eqno(1.6.1)\] 

As we can seen from (1.6.1), variables $\bsym{x},\bsym{p}$ for translational degrees of freedom
are not subjected to transformations, they play a passive role
in the reduction process and they can be omited in further discussion.

In general case kinetic energy of the rigid body is the left-invariant,
but not the right-invariant.
However, for the symmetric top, the system is invariant concerning the {\it right} translation from $S^1 \in SO(3)$,
where group $S^1$ of body symmetry (it is assumed that the axis of body symmetry is directed along vector $\bsym{E}_3$).

As in {\it Section 2} the conditions of {\it Theorem 10.5.1} \cite[p. 355]{MarRat98} are satisfied,
and therefore, we get
\[
   P_1 = T^*SE(3)/ S^1 \simeq T^*\mathbb{R}^3\times K_1 \subset P = T^*\mathbb{R}^3\times K.
\label{eq22}
\eqno(1.6.2)\]
where $P_1$ is the Poisson manifold with such Poisson brackets
\[
\{x_i, p_j\} = \delta_{ij}, \quad
   \{\nu_i, \nu_k\} = 0, \quad
   \{\pi_i, \nu_j\} = \varepsilon_{ijl}\nu_l, \quad
   \{\pi_i, \pi_j\} = \varepsilon_{ijl}\pi_l.
\label{eq23}
\eqno(1.6.3)\]

In order to realize a system reduction completely,
it is necessary to transform the standard Hamiltonian for a rigid body into inertial system,
namely, kinetic energy contribution of proper body rotation.
This is not difficult to perform for the symmetric top,
where two momenta of inertia are equal in the body frame.

Using $I_1 = I_2 = I_\perp$, after a few transformations we obtain
\[
   T_{spin}(((\bsym{x},\bsym{p}),(\mathbf{R},\bsym{\pi})))
   = \frac1{2 I_1} \bsym{\pi}^2
   + \left(\frac1{2 I_3} - \frac1{2 I_1}\right)\langle\bsym{\nu}, \bsym{\pi}\rangle^2
   + V(\bsym{x},\bsym{\nu}).
\label{eq24}
\eqno(1.6.4)\]

That means that the Hamiltonian of the symmetric top takes the form
\[
   h(((\bsym{x},\bsym{p}),(\bsym{\nu},\bsym{\pi})) )
   = \frac1{2 M} \bsym{p}^2 + \frac1{2 I_1} \bsym{\pi}^2
   + V(\bsym{x},\bsym{\nu}),
\label{eq25}
\eqno(1.6.5)\]
after discarding Casimir function
$\left(\frac1{2 I_3} - \frac1{2 I_1}\right)\langle\bsym{\nu}, \bsym{\pi}\rangle^2$.
Where $M$ is a body mass.

Hamiltonian $h$ depends only on the dynamic variables on $P_1$.

The dynamic system is finally reduced to $(P_1,\{\cdot,\cdot\},h)$,
as according to {\it Theorem 10.5.1} \cite[p. 355]{MarRat98}
the dynamic trajectories of the original system are projected onto
the dynamic trajectories of the reduced system by Poisson mapping
\[
   T^*SE(3)\ni ((\bsym{x},\bsym{p}),(\mathbf{R},\bsym{\pi}))
   \mapsto ((\bsym{x},\bsym{p}),(\tau(\mathbf{R}),\bsym{\pi}))
   \in P_1.
\label{eq26}
\eqno(1.6.6)\]

As for the structure of the symplectic of the symplectic leaves
$\textsl{L}^{P_1}_{(\bsym{\nu}_0,\bsym{\pi}_0)}$ of the Poisson manifold $P_1$,
taking the results of {\it Section 4} (see {\bf Proposition 3}) into consideration, we have
\[
   \textsl{L}^{P_1}_{(\bsym{\nu}_0,\bsym{\pi}_0)}
   = T^*\mathbb{R}^3\times \textsl{O}_{(\bsym{\nu}_0,\bsym{\pi}_0)}
\label{D11}
\eqno(1.6.7)\]

\section{ Momentum map for the action of group $SE(3)$}
\label{MomMapSE3}

\bigskip
\subsection{ Momentum map for the action of group $SE(3)$ on itself by left translations}

Let's consider the relations that are important for understanding a momentum map for group $SE(3)$
and for understanding representation of the inertial system.

Let's give the invariant form for Poisson brackets in the inertial system.

Let's assume that $\bsym{v}$ and $\bsym{\omega}$ are the constant arithmetic vectors,
i.e. their components are constant in the inertial system.

We have
\[
   \{x_i, p_j\} = \delta_{ij}, \quad
   \{\pi_i, \pi_j\} = \varepsilon_{ijl}\pi_l, \quad
   \{\pi_i, R_{jk}\} = \varepsilon_{ijl}R_{lk}
\label{34}
\eqno(1.7.1.1)\]

Hence
\[
\{\bsym{x}, \langle \bsym{p},\bsym{v}\rangle\} = \bsym{v}
\label{35}
\eqno(1.7.1.2)\]

We also have
\[
\{\mathbf{R}, \langle \bsym{\pi},\bsym{\omega}\rangle\} = \widehat{\bsym{\omega}}\mathbf{R}
\label{36}
\eqno(1.7.1.3)\] 

Similarly, we have
\[
\{\bsym{\pi}, \langle \bsym{\pi},\bsym{\omega}\rangle\}
   = \bsym{\omega}\times\bsym{\pi} = \widehat{\bsym{\omega}}\bsym{\pi}
\label{37}
\eqno(1.7.1.4)\] 

The relations (1.7.1.3) and (1.7.1.4) can also be written in the form using physical vectors
\[
\{\vec{E}_{k}, \langle \bsym{\pi},\bsym{\omega}\rangle\}
   = \vec{\omega}\times\vec{E}_{k},\quad \vec{\omega} = \omega_i\vec{e}_i
\label{38}
\eqno(1.7.1.5)\] 
\[
\{\vec{\pi}, \langle \bsym{\pi},\bsym{\omega}\rangle\}
   = \vec{\omega}\times\vec{\pi}
\label{39}
\eqno(1.7.1.6)\] 

Expressions (1.7.1.5),(1.7.1.6) show that $\bsym{\pi}$ is the angular momentum of self-rotation of the body
as it generates only rotation of the vectors describing self-rotation of the rigid body.

To generate the rotation of the all physical vectors of the system we should take the total angular momentum of the system
(i.e. including the orbital angular momentum).
\[
\bsym{j} = \bsym{\pi} + \bsym{x}\times\bsym{p}
\label{40}
\eqno(1.7.1.7)\] 

Then we get
\[
   \begin{cases}
      \{\vec{x}, \langle\bsym{j}, \bsym{\omega}\rangle\} = \vec{\omega}\times\vec{x};\\
      \{\vec{p}, \langle\bsym{j}, \bsym{\omega}\rangle\} = \vec{\omega}\times\vec{p};\\
      \{\vec{E}_k, \langle\bsym{j}, \bsym{\omega}\rangle\} = \vec{\omega}\times\vec{E}_k;\\
      \{\vec{\pi}, \langle\bsym{j}, \bsym{\omega}\rangle\} = \vec{\omega}\times\vec{\pi};\\
   \end{cases}
\label{41}
\eqno(1.7.1.8)\] 

From equetions (1.7.1.8) and (1.7.1.2) we find the momentum map that corresponds to the action of group $SE(3)$
on itself by {\it left} translations.
\[
    \bsym{J}_L: ((\bsym{x},\bsym{p}),(\mathbf{R},\bsym{\pi}))
    \mapsto (\bsym{p},\bsym{\pi} + \bsym{x}\times\bsym{p})
\label{42}
\eqno(1.7.1.9)\] 

\bigskip
\subsection{ Momentum map for the action of group $SO(3)$ on $SE(3)$ by right translations}

Let's find the momentum map corresponding to the right action of group $SO(3)$ on $SE(3)$.

Let $\bsym{\Omega}$ be the constant arithmetic vector, that is $\vec{\omega}$
has the constant components in the body frame.

Then
\[
\{\mathbf{R}, \langle \bsym{\Pi},\bsym{\Omega}\rangle\} = \mathbf{R}\widehat{\bsym{\Omega}}
\label{43}
\eqno(1.7.2.1)\] 
\[
\{\vec{E}_{k}, \langle \bsym{\Pi},\bsym{\Omega}\rangle\}
   = \vec{\omega}\times\vec{E}_{k},\quad \vec{\omega} = \Omega_i\vec{E}_i
\label{44}
\eqno(1.7.2.2)\] 

Despite the similarity with formula (1.7.1.4) the Hamiltonian $\langle\bsym{\Pi},\bsym{\Omega}\rangle$
generates such a rotation that the angular velocity in the rigid body, not in the inertial system, is constant.
In this case the angular velocity vector is undergoes rotation: $\vec{\omega}=R_{ik}\Omega^k\vec{e}_i$.

Also we have
\[ \{\bsym{\Pi}, \langle \bsym{\Pi},\bsym{\Omega}\rangle\}
   = -\bsym{\Omega}\times \bsym{\Pi} = \bsym{\Pi}^T\widehat{\bsym{\Omega}}
\label{45}
\eqno(1.7.2.3)\] 

It is in this way that the arithmetic vector $\bsym{\Pi}$ should be changed
in oder for a physical vector $\vec{\pi}=\pi_k\vec{e}_{k}=\Pi_k\vec{E}_{k}$ to get
\[
\{\vec{\pi}, \langle \bsym{\Pi},\bsym{\Omega}\rangle\} = 0
\label{46}
\eqno(1.7.2.4)\] 

Equations (1.7.2.1),(1.7.2.4) show that momentum map corresponding to the right action of group $SO(3)$ on $SE(3)$ is
\[
\bsym{J}_R = \bsym{\Pi}
\label{47}
\eqno(1.7.2.5)\] 

For a general top the kinetic energy is the left-, but not the right-invariant function,
so the components of this moment will not conserved.
In the case of the symmetric top of component $\Pi_3$ is conserved,
and in case of reduction (1.6.2) to the symmetric top this component transfers into the Casimir function $C_2$
in accordance with the general rule (see 12.6.1 \cite[pp. 421-422]{MarRat98}).

\bigskip
\subsection{ Momentum map for the action of group $SE(3)$ on a Poisson manifold P}

For left translation on $T^*SE(3)$ in the inertial system we have
\[
  L^{ct}_{(\bsym{a},\mathbf{A})}: ((\bsym{x},\bsym{p}),(\mathbf{R},\bsym{\pi}))
      \rightarrow((\bsym{a}+\mathbf{A}\bsym{x},\mathbf{A}\bsym{p}),
      (\mathbf{A}\mathbf{R}, \mathbf{A}\bsym{\pi}))\\
\label{48}
\eqno(1.7.3.1)\] 

As
\[
  \tau: \mathbf{R}\mapsto \bsym{\nu} = R_{i3}\bsym{e}_i, \quad \nu_i = R_{i3}
\label{49}
\eqno(1.7.3.2)\] 
then
\[
  \tau(\mathbf{B}\mathbf{R})_{3}= (\mathbf{B}\mathbf{R})_{i3}
   = B_{ik}R_{k3} = B_{ik}\nu_k
\label{50}
\eqno(1.7.3.3)\] 
that means
\[
  \tau(\mathbf{B}\mathbf{R})= \mathbf{B}[\bsym{\nu}]
\label{51}
\eqno(1.7.3.4)\] 

Consequently, we have for the left action $SO(3)$ on $P$
\[
   l_{(\bsym{a},\mathbf{A})}((\bsym{x},\bsym{p}),(\bsym{\nu},\bsym{\pi}))
   = ((\bsym{a}+\mathbf{A}\bsym{x},\mathbf{A}\bsym{p}),(\mathbf{A}\bsym{\nu},\mathbf{A}\bsym{\pi}))
\label{52}
\eqno(1.7.3.5)\] 

Momentum map for action (1.7.3.5) looks like (1.7.1.9)
\[
   \bsym{J}_L: ((\bsym{x},\bsym{p}),(\bsym{\nu},\bsym{\pi}))
   \mapsto (\bsym{p},\bsym{\pi} + \bsym{x}\times\bsym{p})
\label{53}
\eqno(1.7.3.6)\] 

\bigskip
\subsection{ Momentum map on a symplectic leaf $\Lambda_{(\bsym{\nu}_0,\bsym{\pi}_0)}$}

First of all, the Casimir functions $C_1,C_2$ are invariant with respect to the action (1.7.3.6)
of the previous section, and, consequently, they conserved symplectic leaves,
and the generators of this action will be tangential to these leaves.

Therefore we should compare the momentum map on the manifold $P$ with its Poisson structure
with the momentum map on the symplectic leaf with the inner Poisson structure of the leaf
that corresponds to the KKS-symplectic structure.

Let $\xi=(\bsym{v},\bsym{\omega})\in \bsym{se(3)}$.
The vector field $\xi_{P}$ has the form
\[
   \xi_{K}((\bsym{x},\bsym{p}),(\bsym{\nu},\bsym{\pi}))
   = ((\bsym{v}+\bsym{\omega}\times\bsym{x},\bsym{\omega}\times\bsym{p}),
      (\bsym{\omega}\times\bsym{\nu},\bsym{\omega}\times\bsym{\pi}))
\label{54}
\eqno(1.7.4.1)\] 

Momentum map is determined from the relations (see (11.2.3), \cite[p. 374]{MarRat98})
\[ \{F,J(\xi)\}  = \partial_{\xi_{K}}F
\label{55}
\eqno(1.7.4.2)\] 

Let's consider the inner Poisson structure on a symplectic leaf (1.6.7).
If $f$ function on the manifold $P$, then we can denote $\bar{f}$ contraction $f$ on submanifold (1.3.2),
that is $\bar{f}=f_{|\Lambda_{(\bsym{\nu}_0,\bsym{\pi}_0)}}$.
Then according to {\bf Proposition~10.4.2} \cite[p. 346]{MarRat98}
$\{\bar{f},\bar{g}\}=\overline{\{f,g\}}$ for example
\[
   \{x_i, p_j\} = \delta_{ij}, \quad
   \{\bar{\nu}_i, \bar{\nu}_k\} = 0, \quad
   \{\bar{\pi}_i, \bar{\nu}_j\} = \varepsilon_{ijl}\bar{\nu}_l, \quad
   \{\bar{\pi}_i, \bar{\pi}_j\} = \varepsilon_{ijl}\bar{\pi}_l.
\label{56}
\eqno(1.7.4.3)\] 

Using basic dynamic variables $x^i,p_i,\bar{\nu}^i,\bar{\pi}_i$ and Poisson brackets (1.7.4.3) as $\bar{F}$
in the expression (1.7.4.2), we find that dynamic variables $\overline{J(\xi)}$ satisfies (1.7.1.8),
that means it is moment for a symplectic structure on $\Lambda_{(\bsym{\nu}_0,\bsym{\pi}_0)}$
and the following statement is true.

\bigskip
\noindent
{\bf Proposition 4}. {\it Let $J:P \rightarrow \bsym{se(3)}^*$ is a momentum map,
that corresponds to the action of group $SE(3)$ on the Poisson manifold~$P$,
and $J_{(\bsym{\nu}_0,\bsym{\pi}_0)}$:$\Lambda_{(\bsym{\nu}_0,\bsym{\pi}_0)}\rightarrow \bsym{se(3)}^*$
is a momentum map, that corresponds to the action of group $SE(3)$ on the symplectic leaf
$\Lambda_{(\bsym{\nu}_0,\bsym{\pi}_0)}$ then

1. $\bsym{J}_L: ((\bsym{x},\bsym{p}),(\mathbf{R},\bsym{\pi}))
    \mapsto (\bsym{p},\bsym{\pi} + \bsym{x}\times\bsym{p}) $.

2. $J_{\Lambda(\bsym{\nu}_0,\bsym{\pi}_0)}= J_{ | \Lambda_{(\bsym{\nu}_0,\bsym{\pi}_0)}}$
}

\section{ The motion equations of a symmetric top in the external field}
\label{ EqMotion }
We have the Poisson brackets
\[
   \{x_i, p_j\} = \delta_{ij}, \quad
   \{\nu_i, \nu_k\} = 0, \quad
   \{\pi_i, \nu_j\} = \varepsilon_{ijl}\nu_l, \quad
   \{\pi_i, \pi_j\} = \varepsilon_{ijl}\pi_l
\label{57}
\eqno(1.8.1)\] 
and Hamiltonian
\[
   h(((\bsym{x},\bsym{p}),(\bsym{\nu},\bsym{\pi})) )
   = \frac1{2 M} \bsym{p}^2 + \frac1{2 I_1} \bsym{\pi}^2
   + V(\bsym{x},\bsym{\nu})
\label{58}
\eqno(1.8.2)\] 

Applying $\dot{f}=\{f,h\}$ to basic dynamic variables, we get
\[
   \begin{cases}
      \dot{\bsym{x}} = \frac1{M} \bsym{p};\\
      \dot{\bsym{p}} = -\nabla^x V(\bsym{x},\bsym{\nu});\\
      \dot{\bsym{\nu}} =\frac1{I_\perp}\bsym{\pi}\times\bsym{\nu};\\
      \dot{\bsym{\pi}} = \nabla^\nu V(\bsym{x},\bsym{\nu})\times\bsym{\nu};\\
   \end{cases}
\label{59}
\eqno(1.8.3)\] 

\section{ Correlation between energy-momentum method and {\it Ratiu-Ortega Theorem} (see theorem 4.8 \cite{RatOrt99} in the problem of Orbitron }
\label{EMvsROT}

\bigskip
\subsection{ Necessary condition}

The Energy-momentum method as well as Ratiu-Ortega Theorem formulate the conditions in a fixed
(reference) point of the suggested relative equilibrium.

Let's add the variables $(\bsym{x},\bsym{p})$ to the list of variables {\bf Proposition 1, item 3}
changing their names to $u_5,u_6,\dots$.
This order and name of variables will be more convenient in the following calculations
without changing anything in the essence.
It is important that variables $c^1,c^2$ <<numerate>> the symplectic layers $\Lambda_{(\bsym{\nu}_0,\bsym{\pi}_0)}$,
while variables $u_1,u_2,\dots$ are the internal variables on the leaf.

So, let $z_0\in \Lambda_{(\bsym{\nu}_0,\bsym{\pi}_0)}\subset P, (\bsym{\nu}_0^2>0)$.
Then there are 2 coordinate systems in the vicinity of this point:
the 1st one is global $z^i=((\bsym{\nu},\bsym{\pi}),(\bsym{x},\bsym{p}))$ and
the 2nd, generally speaking, is a local $(c^1,c^2,u^1,u^2,u^3,u^4,\dots)$.

We assume that the {\it augmented} Hamiltonian $\bar{h}^\xi$
on the symplectic leaf is the restriction of the augmented Hamiltonian $h^\xi$ on $P$
(this is done for the Orbitron and in a more general case, for example,
for polynomial according to $(\bsym{\nu},\bsym{\pi}))$ Hamiltonians).

So, the 1st condition in the Energy-momentum method is
\[
  {\rm d}\bar{h}^\xi_{|z_0} = 0
\label{60}
\eqno(1.9.1.1)\] 

For any function $f$ in the vicinity $z_0$ we have
\[
  {\rm d}f = {\rm d}_c f + {\rm d}_u f
\label{61}
\eqno(1.9.1.2)\] 

Condition (1.9.1.1) can be written down as
\[
  {\rm d}_u h^\xi_{|z_0} = 0
\label{62}
\eqno(1.9.1.3)\] 

It is also obvious that
\[
   \begin{cases}
      {\rm d}_u C_1 = 0;\\
      {\rm d}_u C_2 = 0;\\
   \end{cases}
\label{63}
\eqno(1.9.1.4)\] 

For any constants $\lambda^1,\lambda^2$ is performed
\[
  {\rm d}_u (h^\xi + \lambda^1 C_1 + \lambda^2 C_2)_{|z_0} = 0
\label{64}
\eqno(1.9.1.5)\] 

In order to fulfill condition (1.9.1.5) not only for {\it partial} but also for the {\it total} differential,
$\lambda^1,\lambda^2$ must have specific definite values that satisfies the equations in point $z_0$,
\[
   \begin{cases}
      \lambda^1\partial_{c^1}C_1 + \lambda^2\partial_{c^1}C_2 = -\partial_{c^1}h^\xi;\\
      \lambda^1\partial_{c^2}C_1 + \lambda^2\partial_{c^2}C_2 = -\partial_{c^2}h^\xi;\\
   \end{cases}
\label{65}
\eqno(1.9.1.6)\] 

Due the special choice of the coordinate system $(c^1,c^2,u^1,u^2,u^3,u^4,\dots)$ we have
\[
   \begin{cases}
      \lambda^1 = -\partial_{c^1}h^\xi;\\
      \lambda^2 = -\partial_{c^2}h^\xi;\\
   \end{cases}
\label{66}
\eqno(1.9.1.7)\] 

If $\lambda^1, \lambda^2$ satisfy (1.9.1.6),(1.9.1.7) then
\[
  {\rm d}h^{\xi,\lambda}_{|z_0} = 0
\label{67}
\eqno(1.9.1.8)\] 
where
\[
  h^{\xi,\lambda} =  h^{\xi} + \lambda^1 C_1 + \lambda^2 C_2
\label{68}
\eqno(1.9.1.9)\] 

Condition (1.9.1.8) has the form of 1st condition in Ratiu-Ortega Theorem.

\bigskip
\subsection{ Sufficient conditions}

Let's consider the 2nd condition of $G_\mu$-stability for Energy-momentum method and Ratiu-Ortega Theorem.

First of all, let's consider Hessian transformation in transition from one local coordinate system ($y^k$)
into another one ($z^i$).
\[
   \frac{\partial^2 f}{\partial z^i\partial z^j}
 = \frac{\partial^2 f}{\partial y^k\partial y^l}\frac{\partial y^k}{\partial z^i}\frac{\partial y^l}{\partial z^j}
 + \frac{\partial f}{\partial y^k}\frac{\partial^2 y^k}{\partial z^i\partial z^j}
\label{69}
\eqno(1.9.2.1)\] 

Hence
\[
  {\rm d}_z^2 f(\eta_1,\eta_2) = \frac{\partial^2 f}{\partial y^k\partial y^l}\nabla_{\eta_1}y^k\nabla_{\eta_2}y^l
  + \frac{\partial f}{\partial y^k}{\rm d}_z^2 y^k(\eta_1,\eta_2),\quad \eta_1,\eta_2\in T_{z_0}P
\label{70}
\eqno(1.9.2.2)\] 

If the coordinate system $y^k$ means $(c^1,c^2,u^1,u^2,u^3,u^4,\dots)$ then
\[
  {\rm d}_z^2 f(w_1,w_2) = \frac{\partial^2 f}{\partial c^A\partial c^B}\nabla_{\eta_1}c^A\nabla_{\eta_2}c^B
\label{71}
\eqno(1.9.2.3)\] 
\[ + \frac{\partial^2 f}{\partial c^A\partial u^\alpha}
     \left(\nabla_{\eta_1}c^A\nabla_{\eta_2}u^\alpha + \nabla_{\eta_2}c^A\nabla_{\eta_1}u^\alpha\right)
\]
\[ + \frac{\partial^2 f}{\partial u^\alpha\partial u^\beta}\nabla_{\eta_1}u^\alpha\nabla_{\eta_2}u^\beta
\]
\[  + \frac{\partial f}{\partial y^k}{\rm d}_z^2 y^k(\eta_1,\eta_2)
\]

Now let's assume that each of the vectors $\eta_1,\eta_2\in T_{z_0}\textsl{L}_{(\bsym{\nu}_0,\bsym{\pi}_0)}$,
that means
\[
    \eta = \eta^\alpha\frac{\partial}{\partial u^\alpha}\longrightarrow
   (\nabla_{\eta}c^A = 0)\&(\nabla_{\eta}u^\alpha = \eta^\alpha)
\label{72}
\eqno(1.9.2.4)\] 
then
\[
  {\rm d}_z^2 f(\eta_1,\eta_2) =
   \frac{\partial^2 f}{\partial u^\alpha\partial u^\beta}\eta^\alpha \eta^\beta
 + \frac{\partial f}{\partial y^k}{\rm d}_z^2 y^k(\eta_1,\eta_2)
\label{73}
\eqno(1.9.2.5)\] 
\[
  {\rm d}_z^2 f(\eta_1,\eta_2) = {\rm d}_u^2 f(\eta_1,\eta_2)
 + \langle\nabla f,{\rm d}_z^2 \bsym{y}(\eta_1,\eta_2)\rangle
\label{74}
\eqno(1.9.2.6)\] 

Let's apply formula (1.9.2.6) to function $h^{\xi,\lambda}$ at the point $z_0$
\[
  {\rm d}_z^2 h^{\xi,\lambda}(\eta_1,\eta_2) = {\rm d}_u^2 h^{\xi,\lambda}(\eta_1,\eta_2)
 + \langle\nabla h^{\xi,\lambda},{\rm d}_z^2 \bsym{y}(\eta_1,\eta_2)\rangle
\label{75}
\eqno(1.9.2.7)\] 

Because of the conditions (1.9.1.8) 2nd term in (1.9.2.7) is equal to 0.
Therefore we have
\[
  {\rm d}_z^2 h^{\xi,\lambda}(\eta_1,\eta_2)
= {\rm d}_u^2 (h^{\xi} + \lambda^1 C_1 + \lambda^2 C_2)(\eta_1,\eta_2)
\label{76}
\eqno(1.9.2.8)\] 
since ${\rm d}_u^2 C_1 = {\rm d}_u^2 C_2 = 0$ then
\[
  {\rm d}_z^2 h^{\xi,\lambda}(\eta_1,\eta_2)
= {\rm d}_u^2 h^{\xi}(\eta_1,\eta_2) = {\rm d}_u^2 \bar{h}^{\xi}(\eta_1,\eta_2).
\label{77}
\eqno(1.9.2.9)\] 

Thus, the positive positive definiteness of the quadratic form in Energy-momentum method (the right side of (1.9.2.9)
is an equivalent to the positive definiteness of the corresponding quadratic form in Ratiu-Ortega Theorem (left side of (1.9.2.9))
on the vectors that are tangent to a symplectic leaf.

We also note that the requirements set by the Energy-momentum method
on the subspace $W$ in a case studied coincide with the corresponding Ratiu-Ortega Theorem.

\newpage
\chapter{Hamiltonian dynamics of a symmetric top in external fields having axial symmetry.\\
Levitating Orbitron}

\section{ Introduction }
\label{intro}

We propose a Hamiltonian that describes a wide class of the
models of a symmetric top that interacts with external field having axial symmetry.
The stability of relative equilibrium is considered.
It investigates on the base on Ratiu-Ortega Theorem.

We found the necessary and sufficient conditions of the dynamic equilibrium.
For the first time in general form the final analytical expressions
for sufficient conditions were deduced.

The general formulas are tested for two important physical models:
the model of the generalized Orbitron (see \cite{GOZ} - without gravity force)
and the model of levitating generalized Orbitron - new task
that generalize the results of the article \cite{ZubLevOrb14}.

Thanks to the new approach became possible to formulate the clear
fully analytical proof that levitating of the Orbitron is possible.

The whole study is conducted in the inertial reference system,
that is justified in the Ch.1 of this article.
Where the equivalence of the algorithms of energy-momentum method
and Ratiu-Ortega Theorem is also proved.

\bigskip

\section{ The motion equations of a symmetric top in an external field}
\label{EqMotion}

We have the Poisson brackets
\[
\{x_i, p_j\} = \delta_{ij}, \quad
   \{\nu_i, \nu_k\} = 0, \quad
   \{\pi_i, \nu_j\} = \varepsilon_{ijl}\nu_l, \quad
   \{\pi_i, \pi_j\} = \varepsilon_{ijl}\pi_l.
\label{A1}
\eqno(2.2.1)\]

\[
   h(((\vec{x}, \vec{\nu}), (\vec{p},\vec{\pi})) )
   = \frac1{2 M} \vec{p}^2 + \frac1{2 I_{\perp}} \vec{\pi}^2
   + V(\vec{x},\vec{\nu}),
\label{A2}
\eqno(2.2.2)\]

Applying $\dot{f}=\{f,h\}$ to the base dynamical variables, we get
\[
   \begin{cases}
      \dot{\vec{x}} = \frac1{M} \vec{p};\\
      \dot{\vec{p}} = -\nabla^x V(\vec{x},\vec{\nu});\\
      \dot{\vec{\nu}} =\frac1{I_\perp}\vec{\pi}\times\vec{\nu};\\
      \dot{\vec{\pi}} = \nabla^\nu V(\vec{x},\vec{\nu})\times\vec{\nu};\\
   \end{cases}
\label{A3}
\eqno(2.2.3)\]

\section{ Axial symmetry condition}
\label{Simmetry}

By virtue of the fact that kinetic energy $SO(3)$-symmetric
then the condition of axial symmetry of the Hamiltonian
can be reduced to the condition of axial symmetry of
the potential energy.
\[
  \{V, j_3\} = x^1\frac{\partial V}{\partial x^2}
             - x^2\frac{\partial V}{\partial x^1}
             + \nu^1\frac{\partial V}{\partial \nu^2}
             - \nu^2\frac{\partial V}{\partial \nu^1} = 0
\label{B1}
\eqno(2.3.1)\]
that is also can be written in the form
\[
  \vec{x}_\perp\times(\nabla^x V)_\perp + \vec{\nu}_\perp\times(\nabla^\nu V)_\perp = 0.
\label{B1a}
\eqno(2.3.1a)\]

In what follows we will show that (2.3.1a) is identically fulfilled
on the orbit of the relative equilibrium, so it bring nothing new
in the study of stability.

\section{ The motion integrals and augmented Hamiltonian}
\label{Augment}

The motion integrals and  augmented Hamiltonian form
remain the same as for the Orbitron, that is
\[
  h^\xi = h - \omega J_3 + \lambda_1 C_1 + \lambda_2 C_2
\label{C1}
\eqno(2.4.1)\]
where
\[
   \begin{cases}
     C_1(\vec{\nu},\vec{\pi}) = \vec{\nu}^2; \\
     C_2(\vec{\nu},\vec{\pi}) = \vec{\nu}\cdot\vec{\pi}; \\
     J_3(((\vec{x}, \vec{\nu}), (\vec{p},\vec{\pi}))) = \pi_3 + x_1 p_2 - x_2 p_1.
   \end{cases}
\label{C2}
\eqno(2.4.2)\]

\section{ Necessary condition of the relative equilibrium}
\label{ Necessary }

\[
  {\rm d}h^\xi = \left(\frac1{M} \vec{p} - \omega\bsym{e}_3\times\vec{x}\right)\cdot{\rm d}\vec{p}
 + \left(\frac1{I_\perp} \vec{\pi} - \omega\vec{e}_3 + \lambda_2\vec{\nu}\right)\cdot{\rm d}\vec{\pi}
\label{D1}
\eqno(2.5.1)\]
\[ + \left(\nabla^x V + \omega\vec{e}_3\times\vec{p}\right)\cdot{\rm d}\vec{x}
   + \left(\nabla^\nu V +  2\lambda_1\vec{\nu} + \lambda_2\vec{\pi}\right)\cdot{\rm d}\vec{\nu}
\]
we getting the necessary condition of the relative equilibrium
\[
   \begin{cases}
      \vec{p} = M\omega(\vec{e}_3\times\vec{x});\\
      \nabla^x V + \omega\vec{e}_3\times\vec{p} = 0;\\
      \vec{\pi} = I_\perp\omega\vec{e}_3 - \lambda_2 I_\perp\vec{\nu};\\
      \nabla^\nu V +  2\lambda_1\vec{\nu} + \lambda_2\vec{\pi} = 0;
   \end{cases}
\label{D2}
\eqno(2.5.2)\]
\[ \begin{cases}
      \nabla^x V = M\omega^2\vec{x}_\perp ;\\
      \lambda_2 = \omega\nu_3 - \frac1{I_\perp}C_2;\\
      \nabla^\nu V +  \lambda\vec{\nu} + I_\perp\omega\lambda_2\vec{e}_3  = 0;
   \end{cases}
\label{D2a}
\eqno(2.5.2a)\]
where
\[ \lambda = 2\lambda_1 - \lambda_2^2I_\perp
\label{D3}
\eqno(2.5.3)\]

It follows
\[
   \begin{cases}
      \nabla^x V = M\omega^2\vec{x}_\perp ;\\
      (\nabla^\nu V)_\perp = -\lambda\vec{\nu}_\perp; \\
      \omega C_2 = \omega\langle\vec{\nu},\vec{\pi}\rangle
      = \frac{\partial V}{\partial\nu_z}
      + (\lambda + I_\perp\omega^2)\nu_z
   \end{cases}
\label{D4}
\eqno(2.5.4)\]

Here, the symbol $\vec{u}_\perp$ is a component of the vector $\vec{u}$
that is orthogonal to the symmetry axis $z$.

{\bf Remark 1}. Vector $\vec{\nu}$ and Lagrange multiplier $\omega$ can be found from first two
lines of (2.5.2) or that is the same from the first line of (2.5.2a). That is distinct
from Orbitron, where we postulate that $\vec{\nu}$ direct in the line of $z$ axis.

{\bf Remark 2}. Formulas (2.5.3) not only determined $\vec{\nu}$ and Lagrange multiplier $\omega,\lambda_1,\lambda_2$,
but most likely applies restrictions on the function $V$ in addition to it axial symmetry.

\section{ Supporting point and allowable variations}
\label{Variation}

\[
   \begin{cases}
      \vec{x}_0 = r_0\vec{e}_1;\\
      \vec{p}_0 = M\omega r_0\vec{e}_2;\\
      \vec{\nu}_0 = \vec{\nu}_0,\quad \vec{\nu}_0^2 = 1; \\
      \vec{\pi}_0 = C_2\vec{\nu}_0 + I_\perp\omega P^{\nu_0}_\perp(\vec{e}_3);\\
   \end{cases}
\label{E1}
\eqno(2.6.1)\]
As it was already mentioned, vector $\vec{\nu}_0$ must be found from the necessary conditions of stability.

Allowable variations, vector $v=(\delta x,\delta p,\delta\nu,\delta \pi)$ in the supporting point fit the conditions of
\[
  \begin{cases}
      \delta_v C_1 = 2\nu_{0i}\delta\nu_i = 0;\\
      \delta_v C_2 = \pi_{0i}\delta\nu_i + \nu_{0i}\delta\pi_i = 0;\\
      \delta_v J_3 = \delta\pi_3 + p_0\delta x_1 + r_0\delta p_2 = 0.\\
   \end{cases}
\label{E2}
\eqno(2.6.2)\]

Adding condition of transversality to these conditions that in contrast
to the Orbitron task we will take in more simpler form that simplifies the further computations.

So, the complete set of the conditions can be written in the following form
\[
   \begin{cases}
      \delta\nu_3 = -\frac{1}{\nu_{0z}}\langle\vec{\nu}_{0\perp},\delta\vec{\nu}\rangle;\\
      \delta\pi_3 =-\frac1{\nu_{0z}}\left\langle\vec{\nu}_{0\perp},\left(\delta\vec{\pi}
      - \frac{I_\perp\omega}{\nu_{0z}}\delta\vec{\nu}\right)\right\rangle;\\
      \delta p_2 = - \frac{p_0}{r_0}\delta x_1 - \frac{1}{r_0}\delta\pi_3 ;\\
      \delta x_2 = 0;
   \end{cases}
\label{E3}
\eqno(2.6.3)\]

{\bf Remark 3}. As it follows from second line of (2.6.1) that for the vector tangent to the orbit the equation $\dot{x}_{02} = 0$
could not be fulfilled in case of conditions of $r_0 \neq 0$ and $\omega \neq 0$ that are always supposed.

\section{ Coordinate system on the plane that is fit to analysis of restraints of the variations}
\label{Coord2}

It is convenient to provide analysis of the restraints of the variations
in the coordinate system that is fit to the vector $\vec{\nu}_{0\perp}$.

{\bf Remark 4}. If this vector is 0 (as in the Orbitron)
then the new coordinate system is coincides with original Cartesian system on the plane.

Let's consider the coordinates system on the plane is defined by the vector of $\vec{a}$.

Let on the plane we defined the coordinate system $\{\vec{e}_1,\vec{e}_2\}$ and vector $\vec{a}$.

Let's direct the basis vector $\vec{E}_1$ in the new coordinate system $\{\vec{E}_1,\vec{E}_2\}$ in the line of vector $\vec{a}$.
In order that new coordinate system has been properly oriented ($\vec{E}_1\times\vec{E}_2=\vec{e}_3$),
the following equations must be satisfied
\[
  \begin{cases}
     \vec{E}_1 = \frac{a^1}{|\vec{a}|}\vec{e}_1 + \frac{a^2}{|\vec{a}|}\vec{e}_2
   = \frac{\vec{a}}{|\vec{a}|};\\
     \vec{E}_2 = -\frac{a^2}{|\vec{a}|}\vec{e}_1 + \frac{a^1}{|\vec{a}|}\vec{e}_2;\\
  \end{cases}
\label{F1}
\eqno(2.7.1)\]

Allow the matrix
\[
  \alpha =
   \begin{bmatrix}
     \frac{a^1}{|\vec{a}|} & -\frac{a^2}{|\vec{a}|}\\
     \frac{a^2}{|\vec{a}|} & \frac{a^1}{|\vec{a}|}\\
   \end{bmatrix},\quad
   \vec{E}_i = \alpha_{k i}\vec{e}_k,\quad
   \alpha_{k i} = \langle \vec{e}_k,\vec{E}_i\rangle
\label{F2}
\eqno(2.7.2)\]

Next
\[  \alpha_{k i}X_i = \langle \vec{e}_k,X_i\vec{E}_i\rangle
    = \langle \vec{e}_k,\vec{x}\rangle = x_k
\]
i.e.
\[  x_k = \alpha_{k i}X_i,\quad \bsym{x} = \alpha\bsym{X}
\label{F2a}
\eqno(2.7.2a)\]
\[
  \begin{cases}
     x_1 = \frac{a_1}{|\vec{a}|}X_1 - \frac{a_2}{|\vec{a}|}X_2;\\
     x_2 = \frac{a_2}{|\vec{a}|}X_1 + \frac{a_1}{|\vec{a}|}X_2;\\
  \end{cases}
\label{F2b}
\eqno(2.7.2b)\]

Correspondingly
\[  X_k = (\alpha^{-1})_{k i}x_i,\quad \bsym{X} = \alpha^{-1}\vec{x}
\label{F3}
\eqno(2.7.3)\]
where
\[ \alpha^{-1} =\alpha^T =
   \begin{bmatrix}
     \frac{a^1}{|\vec{a}|}  & \frac{a^2}{|\vec{a}|}\\
     -\frac{a^2}{|\vec{a}|} & \frac{a^1}{|\vec{a}|}\\
   \end{bmatrix}
\label{F3a}
\eqno(2.7.3a)\]

In our case, the vector defined on the plane is $\vec{\nu}_{0\perp}$, so
\[  \alpha = \frac{1}{|\vec{\nu}_{0\perp}|}
    \begin{bmatrix}
       \nu_{01} & -\nu_{02} \\
       \nu_{02} & \nu_{01} \\
    \end{bmatrix},\quad
    \vec{E}_A = \alpha_{B A}\vec{e}_B,\quad
    \alpha_{B A} = \langle \vec{e}_B,\vec{E}_A\rangle
\label{F4}
\eqno(2.7.4)\]
\[
\begin{cases}
   \vec{E}_1 = \frac{1}{|\vec{\nu}_{0\perp}|}(\nu_{01}\vec{e}_1 + \nu_{02}\vec{e}_2)
   = \frac{\vec{\nu}_{0\perp}}{|\vec{\nu}_{0\perp}|};\\
   \vec{E}_2 = \frac{1}{|\vec{\nu}_{0\perp}|}(-\nu_{02}\vec{e}_1 + \nu_{01}\vec{e}_2);\\
\end{cases}
\label{F5}
\eqno(2.7.5)\]

\section{ Restraints of the variations in the new coordinate system }
\label{Constraint}

We shall use the reference coordinate system for the variations $\delta x,\delta p$
and the basis $\{\vec{E}_1,\vec{E}_2,\vec{e}_3\}$ for the variations $\delta\nu,\delta\pi$,
and denote these variations in the new coordinate system by the capital letters $\delta{\mathrm N},\delta\Pi$.

Then restraints (2.6.3) have the form
\[
   \begin{cases}
      \delta\nu_3 = -\frac{|\vec{\nu}_{0\perp}|}{\nu_{0z}}\delta{\mathrm N}_1;\\
      \delta\pi_3 = -\frac{|\vec{\nu}_{0\perp}|}{\nu_{0z}}\left(\delta\Pi_1
                  - \frac1{\nu_{0z}}I_\perp\omega\delta{\mathrm N}_1\right);\\
      \delta p_2 =  - M\omega\delta x_1 + \frac{|\vec{\nu}_{0\perp}|}{r_0\nu_{0z}}
      \left(\delta\Pi_1 - \frac1{\nu_{0z}}I_\perp\omega\delta{\mathrm N}_1\right);\\
      \delta x_2 = 0;
   \end{cases}
\label{G1}
\eqno(2.8.1)\]

Allow the variable $\delta\Pi'_1$
\[
   \begin{cases}
      \delta\Pi'_1 = \delta\Pi_1 - \frac1{\nu_{0z}}I_\perp\omega\delta{\mathrm N}_1;\\
      \delta\Pi_1 = \delta\Pi'_1 + \frac1{\nu_{0z}}I_\perp\omega\delta{\mathrm N}_1;\\
   \end{cases}
\label{G2}
\eqno(2.8.2)\]

Then
\[
   \begin{cases}
      \delta\nu_3 = -\frac{|\vec{\nu}_{0\perp}|}{\nu_{0z}}\delta{\mathrm N}_1;\\
      \delta\pi_3 = -\frac{|\vec{\nu}_{0\perp}|}{\nu_{0z}}\delta\Pi'_1;\\
      \delta p_2 =  - M\omega\delta x_1 + \frac{|\vec{\nu}_{0\perp}|}{r_0\nu_{0z}}\delta\Pi'_1;\\
      \delta x_2 = 0;
   \end{cases}
\label{G1a}
\eqno(2.8.1a)\]

\section{ The original and reduced quadratic forms }
\label{Quadratic}

The second variation of the augmented Hamiltonian has the form (hereinafter indices $i,j,k=1..3,A,B,C=1..2$)
\[ \delta_v^2 h^\lambda = \frac1{M} \delta p_3^2 + \frac1{M} \delta p_1^2 + \frac1{M} \delta p_2^2
\label{H1}
\eqno(2.9.1)\]
\[  + \frac1{I_\perp} \delta\vec{\pi}_\perp^2
    + \frac1{I_\perp}\delta\pi_3^2
\]
\[ + 2\lambda_2\langle\delta\vec{\pi}_\perp, \delta\vec{\nu}_\perp\rangle
   + 2\lambda_2\delta\pi_3\delta\nu_3
\]
\[ + 2\lambda_1\delta\vec{\nu}_\perp^2 + 2\lambda_1\delta\nu_3^2
\]
\[ - 2\omega (\delta x_1\delta p_2 - \delta x_2\delta p_1 )
   + \frac{\partial^2 V}{\partial x^i\partial x^j}\delta x^i\delta x^j
\]
\[ + 2\frac{\partial^2 V}{\partial x^i\partial \nu^A}\delta x^i\delta \nu^A
   + 2\frac{\partial^2 V}{\partial x^i\partial \nu^3}\delta x^i\delta \nu^3
\]
\[ + \frac{\partial^2 V}{\partial \nu^A\partial \nu^B}\delta \nu^A\delta \nu^B
   + 2\frac{\partial^2 V}{\partial \nu^A\partial \nu^3}\delta \nu^A\delta \nu^3
   + \frac{\partial^2 V}{\partial \nu_3^2}\delta \nu_3^2
\]

By using restraints we get
\[
   \begin{cases}
      \frac1{M} \delta p_2^2
      = M\omega^2\delta x_1^2 - 2\omega\frac{|\vec{\nu}_{0\perp}|}{r_0\nu_{0z}}\delta x_1\delta\Pi'_1
      + \frac{\vec{\nu}_{0\perp}^2}{M r_0^2\nu_{0z}^2}\delta\Pi'^2_1;\\
      \frac1{I_\perp} \delta\vec{\pi}^2
      = \frac1{I_\perp}\frac1{\nu_{0z}^2}\delta\Pi'^2_1
      + 2\frac1{\nu_{0z}}\omega\delta\Pi'_1\delta{\mathrm N}_1
      + \frac1{\nu_{0z}^2}I_\perp\omega^2\delta{\mathrm N}_1^2
      + \frac1{I_\perp}\delta\Pi_2^2;\\
      \langle\delta\vec{\pi}, \delta\vec{\nu}\rangle
      = \frac1{\nu_{0z}^2}\delta\Pi'_1\delta{\mathrm N}_1
      + \delta\Pi_2\delta{\mathrm N}_2
      + \frac1{\nu_{0z}}I_\perp\omega\delta{\mathrm N}_1^2;\\
      \delta\vec{\nu}^2 = \frac1{\nu_{0z}^2}\delta{\mathrm N}_1^2 + \delta{\mathrm N}_2^2;\\
      - 2\omega (\delta x_1\delta p_2 - \delta x_2\delta p_1 )
      = 2 M\omega^2\delta x_1^2 - 2\omega\frac{|\vec{\nu}_{0\perp}|}{r_0\nu_{0z}}\delta x_1\delta\Pi'_1
   \end{cases}
\label{H2}
\eqno(2.9.2)\]

Hereinafter we use the following notation
\[
   \begin{cases}
     \frac{\partial^2 V}{\partial x^i\partial{\mathrm N}_A}
     = {\mathrm d}^2V(\vec{e}_i, \vec{E}_A); \\
     \frac{\partial^2 V}{\partial{\mathrm N}_A\partial{\mathrm N}_B}
     = {\mathrm d}^2V(\vec{E}_A, \vec{E}_B); \\
     \frac{\partial^2 V}{\partial{\mathrm N}_A\partial\nu_3}
     =\partial_{\vec{E}_A}\left(\frac{\partial V}{\partial\nu_3}\right);\\
   \end{cases}
\label{H3}
\eqno(2.9.3)\] 
besides that in the expressions of this type we always considered that the vectors $\vec{e}_i$
are tangent to the variables space of $\vec{x}$, and $\vec{E}_A$ to the variables space of $\vec{\nu}$.

Substituting (2.9.2) into (2.9.1) we find the reduced quadratic form
\[ \delta_v^2 h^\lambda = \frac1{M} \delta p_3^2 + \frac1{M} \delta p_1^2
\label{H4}
\eqno(2.9.4)\] 
\[    + M\omega^2\delta x_1^2 - 2\omega\frac{|\vec{\nu}_{0\perp}|}{r_0\nu_{0z}}\delta x_1\delta\Pi'_1
      + \frac{\vec{\nu}_{0\perp}^2}{M r_0^2\nu_{0z}^2}\delta\Pi'^2_1
\]
\[ + \frac1{I_\perp}\frac1{\nu_{0z}^2}\delta\Pi'^2_1
   + 2\frac1{\nu_{0z}}\omega\delta\Pi'_1\delta{\mathrm N}_1
   + \frac1{\nu_{0z}^2}I_\perp\omega^2\delta{\mathrm N}_1^2
   + \frac1{I_\perp}\delta\Pi_2^2
\]
\[ + 2\lambda_2\left(\frac1{\nu_{0z}^2}\delta\Pi'_1\delta{\mathrm N}_1
      + \delta\Pi_2\delta{\mathrm N}_2
      + \frac1{\nu_{0z}}I_\perp\omega\delta{\mathrm N}_1^2\right)
\]
\[ + 2\lambda_1\left(\frac1{\nu_{0z}^2}\delta{\mathrm N}_1^2 + \delta{\mathrm N}_2^2\right)
\]
\[ + 2 M\omega^2\delta x_1^2 - 2\omega\frac{|\vec{\nu}_{0\perp}|}{r_0\nu_{0z}}\delta x_1\delta\Pi'_1
\]
\[ + \frac{\partial^2 V}{\partial x_1^2}\delta x_1^2
   + 2\frac{\partial^2 V}{\partial x_1\partial x_3}\delta x_1\delta x_3
   + \frac{\partial^2 V}{\partial x_3^2}\delta x_3^2
\]
\[ + 2\frac{\partial^2 V}{\partial x^1\partial{\mathrm N}_A}\delta x^1\delta{\mathrm N}_A
   + 2\frac{\partial^2 V}{\partial x^3\partial{\mathrm N}_A}\delta x^3\delta{\mathrm N}_A
\]
\[ - 2\frac{|\vec{\nu}_{0\perp}|}{\nu_{0z}}\frac{\partial^2 V}{\partial x^1\partial\nu^3}\delta x^1\delta{\mathrm N}_1
   - 2\frac{|\vec{\nu}_{0\perp}|}{\nu_{0z}}\frac{\partial^2 V}{\partial x^3\partial\nu^3}\delta x^3\delta{\mathrm N}_1
\]
\[ + \frac{\partial^2 V}{\partial{\mathrm N}_1^2}\delta{\mathrm N}_1^2
    + 2\frac{\partial^2 V}{\partial{\mathrm N}_1\partial{\mathrm N}_2}
     \delta{\mathrm N}_1\delta{\mathrm N}_2
    + \frac{\partial^2 V}{\partial{\mathrm N}_2^2}\delta{\mathrm N}_2^2
\]
\[ - 2\frac{|\vec{\nu}_{0\perp}|}{\nu_{0z}}\frac{\partial^2 V}{\partial{\mathrm N}_1\partial\nu^3}\delta{\mathrm N}_1^2
   - 2\frac{|\vec{\nu}_{0\perp}|}{\nu_{0z}}\frac{\partial^2 V}{\partial{\mathrm N}_2\partial\nu^3}\delta{\mathrm N}_1\delta{\mathrm N}_2
\]
\[ + \frac{\vec{\nu}_{0\perp}^2}{\nu_{0z}^2}\frac{\partial^2 V}{\partial \nu_3^2}\delta{\mathrm N}_1^2
\]

\section{ The method of successive elimination of isolated squares }
\label{Isolate}

To derive the conditions of positive definiteness of the quadratic form
\[ Q = Q(x_1, \dots, x_n)
\label{I1}
\eqno(2.10.1)\]
we use the method of successive elimination of isolated squares.

Let's represent $Q$ in the form
\[ Q(x_1, \dots, x_n) = A x_1^2 + 2 B (x_2, \dots, x_n) x_1 + Q'(x_2, \dots, x_n),
\label{I2}
\eqno(2.10.2)\]
where $B$ linear function of its variables, and $Q'$ is quadratic function of the variables but without $x_1$.

For the positive definiteness of $Q$ is necessary that $A>0$ then
$Q$ positive definiteness is equivalent to the positive definiteness of the quadratic form $Q_1$,
but with smaller number of variables.
\[ Q_1(x_2, \dots, x_n) = -\frac{1}{A} B^2(x_2, \dots, x_n) + Q'(x_2, \dots, x_n),
\label{I3}
\eqno(2.10.3)\]
Consistently applying this procedure, we can find all the conditions for positive
definiteness of the original quadratic form, that is means all $A>0$.

{\bf Remark 5}. The order of elimination of the variables is arbitrary.

\section{ Sufficient conditions of stability }
\label{Sufficient}

Applying the method of successive elimination of isolated squares in the following order
$\delta p_3,\delta p_1,\delta\Pi_2,\delta\Pi'_1,\delta{\mathrm N}_2,\delta{\mathrm N}_1$, result
\[ \begin{cases}
      \lambda + {\mathrm d}^2V(\vec{E}_2, \vec{E}_2)> 0;\\
      \lambda + {\mathrm d}^2V(\vec{\nu}_0^\top, \vec{\nu}_0^\top)
   + \frac{I_\perp^2\vec{\nu}_{0\perp}^2}{I_\perp\vec{\nu}_{0\perp}^2 + M r_0^2}
     \left(\nu_{0z}\omega + \lambda_2\right)^2
   + \vec{\nu}_{0\perp}^2 I_\perp\omega^2
   -\frac{{\mathrm d}^2V(\vec{E}_2, \vec{\nu}_0^\top)^2}
   {\lambda + {\mathrm d}^2V(\vec{E}_2, \vec{E}_2)} > 0;\\
   A > 0,\quad C > 0,\quad A C - B^2 > 0;\\
   \end{cases}
\label{J1}
\eqno(2.11.1)\]
where
\[  A =
    M\omega^2\frac{3 M r_0^2 - I_\perp\vec{\nu}_{0\perp}^2}{I_\perp\vec{\nu}_{0\perp}^2 + M r_0^2}
    + \frac{\partial^2 V}{\partial x_1^2}
    - \frac{\left(\frac{\partial^2 V}{\partial x_1\partial{\mathrm N}_2}\right)^2}
    {\lambda + \frac{\partial^2 V}{\partial{\mathrm N}_2^2}}
\label{J2a}
\eqno(2.11.2a)\]
\[ -\frac{\left(2\frac{I_\perp|\vec{\nu}_{0\perp}| p_0}{I_\perp\vec{\nu}_{0\perp}^2 + M r_0^2}
      \left(\nu_{0z}\omega + \lambda_2\right)
   + {\mathrm d}^2V(\vec{e}_1, \vec{\nu}_0^\top)
   - \frac{\frac{\partial^2 V}{\partial x^1\partial{\mathrm N}_2}{\mathrm d}^2V(\vec{E}_2, \vec{\nu}_0^\top)}
   {\lambda + \frac{\partial^2 V}{\partial{\mathrm N}_2^2}}\right)^2}
   {\lambda + {\mathrm d}^2V(\vec{\nu}_0^\top, \vec{\nu}_0^\top)
   + \frac{I_\perp^2\vec{\nu}_{0\perp}^2}{I_\perp\vec{\nu}_{0\perp}^2 + M r_0^2}
     \left(\nu_{0z}\omega + \lambda_2\right)^2
   + \vec{\nu}_{0\perp}^2 I_\perp\omega^2
   -\frac{{\mathrm d}^2V(\vec{E}_2, \vec{\nu}_0^\top)^2}
   {\lambda + \frac{\partial^2 V}{\partial{\mathrm N}_2^2}}}
\]
\[  B = \frac{\partial^2 V}{\partial x_1\partial x_3} - \frac{\left(\frac{\partial^2 V}{\partial x_1\partial{\mathrm N}_2}\right)
     \left(\frac{\partial^2 V}{\partial x_3\partial{\mathrm N}_2}\right)}
     {\lambda + \frac{\partial^2 V}{\partial{\mathrm N}_2^2}}
\label{J2b}
\eqno(2.11.2b)\]
\[ - \frac{\left(2\frac{I_\perp|\vec{\nu}_{0\perp}| p_0}{I_\perp\vec{\nu}_{0\perp}^2 + M r_0^2}
      \left(\nu_{0z}\omega + \lambda_2\right)
   + {\mathrm d}^2V(\vec{e}_1, \vec{\nu}_0^\top)
   - \frac{\frac{\partial^2 V}{\partial x^1\partial{\mathrm N}_2}{\mathrm d}^2V(\vec{E}_2, \vec{\nu}_0^\top)}
   {\lambda + \frac{\partial^2 V}{\partial{\mathrm N}_2^2}}\right)}
   {\lambda + {\mathrm d}^2V(\vec{\nu}_0^\top, \vec{\nu}_0^\top)
   + \frac{I_\perp^2\vec{\nu}_{0\perp}^2}{I_\perp\vec{\nu}_{0\perp}^2 + M r_0^2}
     \left(\nu_{0z}\omega + \lambda_2\right)^2
   + \vec{\nu}_{0\perp}^2 I_\perp\omega^2
   -\frac{{\mathrm d}^2V(\vec{E}_2, \vec{\nu}_0^\top)^2}
   {\lambda + \frac{\partial^2 V}{\partial{\mathrm N}_2^2}}}
\]
\[ \times\left({\mathrm d}^2V(\vec{e}_3, \vec{\nu}_0^\top)
   - \frac{\frac{\partial^2 V}{\partial x_3\partial{\mathrm N}_2}{\mathrm d}^2V(\vec{E}_2, \vec{\nu}_0^\top)}
     {\lambda + \frac{\partial^2 V}{\partial{\mathrm N}_2^2}}\right)
\]

\[ C = \frac{\partial^2 V}{\partial x_3^2}
   - \frac{\left(\frac{\partial^2 V}{\partial x_3\partial{\mathrm N}_2}\right)^2}
     {\lambda + \frac{\partial^2 V}{\partial{\mathrm N}_2^2}}
\label{J2c}
\eqno(2.11.2c)\]
\[ - \frac{\left({\mathrm d}^2V(\vec{e}_3, \vec{\nu}_0^\top)
   - \frac{\frac{\partial^2 V}{\partial x_3\partial{\mathrm N}_2}{\mathrm d}^2V(\vec{E}_2, \vec{\nu}_0^\top)}
     {\lambda + \frac{\partial^2 V}{\partial{\mathrm N}_2^2}}\right)^2}
     {\lambda + {\mathrm d}^2V(\vec{\nu}_0^\top, \vec{\nu}_0^\top)
   + \frac{I_\perp^2\vec{\nu}_{0\perp}^2}{I_\perp\vec{\nu}_{0\perp}^2 + M r_0^2}
     \left(\nu_{0z}\omega + \lambda_2\right)^2
   + \vec{\nu}_{0\perp}^2 I_\perp\omega^2
   -\frac{{\mathrm d}^2V(\vec{E}_2, \vec{\nu}_0^\top)^2}
   {\lambda + \frac{\partial^2 V}{\partial{\mathrm N}_2^2}}}
\]

We use the following notation.

Allow the vector
\[  \vec{\nu}_0^\top = \nu_{0z}\vec{E}_1 - |\vec{\nu}_{0\perp}|\vec{e}_3,
    \quad \langle\vec{\nu}_0, \vec{\nu}_0^\top\rangle = 0
\label{J3}
\eqno(2.11.3)\]

Then
\[
   \begin{cases}
      {\mathrm d}^2V(\vec{E}_2, \vec{\nu}_0^\top) =
      \nu_{0z}\frac{\partial^2 V}{\partial{\mathrm N}_1\partial{\mathrm N}_2}
      - |\vec{\nu}_{0\perp}|\frac{\partial^2 V}{\partial{\mathrm N}_2\partial\nu^3};\\
      {\mathrm d}^2V(\vec{e}_1, \vec{\nu}_0^\top) =
      \nu_{0z}\frac{\partial^2 V}{\partial x^1\partial{\mathrm N}_1}
      - |\vec{\nu}_{0\perp}|\frac{\partial^2 V}{\partial x^1\partial\nu^3};\\
      {\mathrm d}^2V(\vec{e}_3, \vec{\nu}_0^\top) =
      \nu_{0z}\frac{\partial^2 V}{\partial x^3\partial{\mathrm N}_1}
      - |\vec{\nu}_{0\perp}|\frac{\partial^2 V}{\partial x^3\partial\nu^3};\\
      {\mathrm d}^2V(\vec{\nu}_0^\top, \vec{\nu}_0^\top) =
      \nu_{0z}^2\frac{\partial^2 V}{\partial{\mathrm N}_1^2}
      - 2|\vec{\nu}_{0\perp}|\nu_{0z}\frac{\partial^2 V}{\partial{\mathrm N}_1\partial\nu^3}
      + \vec{\nu}_{0\perp}^2\frac{\partial^2 V}{\partial \nu_3^2}
   \end{cases}
\label{J4}
\eqno(2.11.4)\]
where
\[
\begin{cases}
   \frac{\partial^2 V}{\partial x_1\partial{\mathrm N}_1}
   = \frac{\partial^2 V}{\partial x_1\partial\nu}_1\frac{\nu_{01}}{|\vec{\nu}_{0\perp}|}
   + \frac{\partial^2 V}{\partial x_1\partial\nu_2}\frac{\nu_{02}}{|\vec{\nu}_{0\perp}|};\\

   \frac{\partial^2 V}{\partial x_1\partial{\mathrm N}_2}
   = -\frac{\partial^2 V}{\partial x_1\partial\nu_1}\frac{\nu_{02}}{|\vec{\nu}_{0\perp}|}
   + \frac{\partial^2 V}{\partial x_1\partial\nu_2}\frac{\nu_{01}}{|\vec{\nu}_{0\perp}|};\\

   \frac{\partial^2 V}{\partial{\mathrm N}_1\partial\nu^3}
   = \frac{\partial^2 V}{\partial\nu_1\partial \nu_3}\frac{\nu_{01}}{|\vec{\nu}_{0\perp}|}
   + \frac{\partial^2 V}{\partial\nu_2\partial \nu_3}\frac{\nu_{02}}{|\vec{\nu}_{0\perp}|};\\

   \frac{\partial^2 V}{\partial{\mathrm N}_2\partial\nu^3}
   = -\frac{\partial^2 V}{\partial\nu_1\partial\nu_3}\frac{\nu_{02}}{|\vec{\nu}_{0\perp}|}
   + \frac{\partial^2 V}{\partial\nu_2\partial\nu_3}\frac{\nu_{01}}{|\vec{\nu}_{0\perp}|};\\

   \frac{\partial^2 V}{\partial{\mathrm N}_1^2}
   = \frac1{|\vec{\nu}_{0\perp}|^2}\left(
   \frac{\partial^2 V}{\partial \nu_1^2} \nu_{01}^2
   + 2\frac{\partial^2 V}{\partial \nu_1\partial\nu_2}\nu_{01}\nu_{02}
   + \frac{\partial^2 V}{\partial \nu_2^2} \nu_{02}^2\right);\\

   \frac{\partial^2 V}{\partial{\mathrm N}_1\partial{\mathrm N}_2}
   = \frac1{|\vec{\nu}_{0\perp}|^2}\left(-\frac{\partial^2 V}{\partial\nu_1^2}\nu_{01}\nu_{02}
   + \frac{\partial^2 V}{\partial \nu_2^2}\nu_{01}\nu_{02}
   + \frac{\partial^2 V}{\partial\nu_1\partial\nu_2}(\nu_{01}^2 - \nu_{02}^2)\right);\\

   \frac{\partial^2 V}{\partial{\mathrm N}_2^2}
   = \frac1{|\vec{\nu}_{0\perp}|^2}\left(
   \frac{\partial^2 V}{\partial \nu_1^2} \nu_{02}^2
   - 2\frac{\partial^2 V}{\partial \nu_1\partial\nu_2}\nu_{01}\nu_{02}
   + \frac{\partial^2 V}{\partial \nu_2^2} \nu_{01}^2\right);\\
\end{cases}
\label{J5}
\eqno(2.11.5)\]

\section{ The case of a dipole in a magnetic field }
\label{Dipol}

Let's study the possibility of the levitation of a dipole
in an axially symmetric magnetic field.

The potential energy in this case has the form
\[ V = -\mu \langle\vec{\nu},\vec{B}\rangle + M g z
   = -\mu\nu^i B_i(\vec{x})  + M g z
\label{K1}
\eqno(2.12.1)\]
where $\vec{B}$ is an axially symmetric magnetic field.
Next, everywhere we assume that $\mu>0$.

From (2.12.1) follows
\[ \nabla^\nu V = -\mu\vec{B}
\label{K2}
\eqno(2.12.2)\]

Hence
\[
   \begin{cases}
      \lambda_2 = \omega\nu_z - \frac1{I_\perp}C_2;\\
      \omega C_2 = - \mu B_z + (\lambda + I_\perp\omega^2)\nu_z;\\
      \vec{\nu}_\perp = \frac{\mu}{\lambda}\vec{B}_\perp;\\
   \end{cases}
\label{K3}
\eqno(2.12.3)\]

From (3) it follows that $\vec{\nu}_\perp$ has the same direction as the $\vec{B}_\perp$ (with the sign).
As it will be shown below, this condition is coherent with the equation
\[ \nabla^x V = M\omega^2\vec{x}_\perp
\label{K4}
\eqno(2.12.4)\]
for the potential energy of the dipole (2.12.1).

Therefore, the supporting point in this case can be selected in form
\[
   \begin{cases}
      \vec{x}_0 = r_0\vec{e}_1;\\
      \vec{p}_0 = M\omega r_0\vec{e}_2;\\
      \vec{\nu}_0 = \nu_{z}\vec{e}_3 + \nu_{r}\vec{e}_1,
      \quad \nu_r^2 + \nu_z^2 =1,
      \quad \nu_r = \frac{\mu}{\lambda}B_r; \\
      \vec{\pi}_0 = I_\perp\omega\vec{e}_3 - \lambda_2 I_\perp\vec{\nu}_0; \\
   \end{cases}
\label{K5}
\eqno(2.12.5)\]
i.e. vector $\vec{\nu}_0$ lies in the plane ${\rm span}(\vec{x}_0,\vec{e}_3)$.

In the dipole case the significant simplification will be
\[ {\mathrm d}^2_\nu V = 0
\label{K6}
\eqno(2.12.6)\]

\subsection{ The relations for axially symmetric magnetic field }

It is assumed that the sources of magnetic field are localized,
and motion of the magnetized body occurs in an area free of sources.

Then in the movement area the equations of the magnetostatics without sources are fulfilled.
For the axially symmetric magnetic field in particular
that means in the cylindrical coordinate system there are only two components of the field $B_r,B_z$
that depend on 2 variables --- $r$ and $z$.

In given case we have the following equations of magnetostatics
\[
  \begin{cases}
     B_{r,z} - B_{z,r} = 0; \\
     B_{z,z} + B_{r,r} + \frac1r B_r = 0;\\
     B_{z,zz} + B_{z,rr} + \frac1r B_{z,r} = 0
  \end{cases}
\label{L1}
\eqno(2.12.1.1)\]

For the Jacobian we have
\[
  \begin{cases}
     B_{A,C} = B_{C,A} = \frac{B_r}{r}\delta_{A C}
     + \left(B_{r,r} - \frac{B_r}{r}\right)\frac{x_A x_C}{r^2}; \\
     B_{3,C} = B_{z,r}\frac{x_C}{r};\\
     B_{3,3} = B_{z,z}
  \end{cases}
\label{L2}
\eqno(2.12.1.2)\]

For the Hessian of the magnetic field we have expressions
\[
  \begin{cases}
     B_{A,CD} = -B_{z,rz}\frac{x_A x_C x_D}{r^3}\\
     - \frac{1}{r}\left(B_{z,z} + \frac{2 B_r}{r}\right)\left(\frac{x_A\delta_{CD} + x_C\delta_{AD} + x_D\delta_{A C} }{r}
     - \frac{4 x_A x_C x_D}{r^3}\right); \\
     B_{3,CD} = \frac{1}{r}B_{z,r}\delta_{CD}
     + \left(B_{z,rr}
     - \frac{1}{r}B_{z,r}\right)\frac{x_C x_D}{r^2};\\
     B_{3,C3} = \frac{x_C}{r}(B_{z,zr});\\
     B_{3,33} = B_{z,zz};
\end{cases}
\label{L3}
\eqno(2.12.1.3)\]

In the supporting point where $\vec{x}=(r_0,0,0)^T$ for the components of the Jacobian we have
\[
  \begin{cases}
     B_{1,1} = B_{r,r}; \\
     B_{1,2} = B_{2,1} = 0; \\
     B_{2,2} =  \frac1r B_r; \\
     B_{3,1} = B_{z,r};\\
     B_{3,2} = 0;\\
     B_{3,3} = B_{z,z}
  \end{cases}
\label{L2a}
\eqno(2.12.1.2a)\]

For the components of the Hessian in the supporting point we have
\[
   \begin{cases}
     B_{1,11} = -B_{z,zr} + \frac1{r}\left(B_{z,z} + \frac2{r}B_r\right)
     = -B_{z,zr} - \frac1{r}\left(B_{r,r} - \frac1{r}B_r\right);\\
     B_{1,12} = B_{1,21} = 0;\\
     B_{1,22} = \frac1{r}\left(B_{z,z} + \frac2{r}B_r\right)
     =  - \frac1{r}\left(B_{r,r} - \frac1{r}B_r\right);\\
     B_{2,11} = B_{1,12} = 0;\\
     B_{2,12} = B_{2,21} = B_{1,22} = \frac1{r}\left(B_{z,z} + \frac2{r}B_r\right)
     =  - \frac1{r}\left(B_{r,r} - \frac1{r}B_r\right);\\
     B_{2,22} = 0;\\
     B_{3,11} = B_{z,rr};\\
     B_{3,12} = 0;\\
     B_{3,13} = B_{z,rz};\\
     B_{3,23} = 0;\\
     B_{3,33} = B_{z,zz};\\
   \end{cases}
\label{L3a}
\eqno(2.12.1.3a)\]

\subsection{ The equations of equilibrium for the magnetic dipole }

From the relation (2.12.1) we have
\[
   \begin{cases}
      V_{,C} = \frac{\partial V}{\partial x^C}
      = -\mu\nu^i B_{i,C} = -\mu\nu^D B_{D,C} - \mu\nu^3 B_{3,C};\\
      V_{,3} = \frac{\partial V}{\partial x^3}
      = -\mu\nu^i B_{i,3} + M g = -\mu\nu^D B_{D,3} - \mu\nu^3 B_{3,3} + M g;\\
   \end{cases}
\label{M1}
\eqno(2.12.2.1)\]

Taking in to account the third equation of (2.12.3) the properties of the axially-symmetric magnetic field we obtain
\[ \begin{cases}
      \nu_r = \frac{\mu}{\lambda}B_r;\\
      \nu^C = \nu^r\frac{x^C}{r};\\
   \end{cases}
\label{M2}
\eqno(2.12.2.2)\]
and equations of equilibrium take the form
\[ \begin{cases}
      \nu^z B_{z,z} + \nu^r B_{r,z}  = \frac{M}{\mu}g;\\
      \nu^z B_{r,z} + \nu^r B_{r,r}  = -\frac{M}{\mu}\omega^2 r;\\
      \nu_r^2 + \nu_z^2 = 1
   \end{cases}
\label{M3}
\eqno(2.12.2.3)\]

\subsection{ The sufficient conditions of stability for the magnetic dipole in an axially symmetric magnetic field }

In the case of the magnetic dipole in the supporting point we have the following relations
\[
  \begin{cases}
     \frac{\partial^2 V}{\partial x^1\partial{\mathrm N}_1}
     = {\rm sign}(\nu_{r})\frac{\partial^2 V}{\partial x_1\partial\nu_1};\\

     \frac{\partial^2 V}{\partial x_1\partial{\mathrm N}_2}
     = {\rm sign}(\nu_{r})\frac{\partial^2 V}{\partial x_1\partial\nu_2};\\

     \frac{\partial^2 V}{\partial{\mathrm N}_1\partial\nu^3}
     = {\rm sign}(\nu_{r})\frac{\partial^2 V}{\partial\nu_1\partial \nu_3};\\

     \frac{\partial^2 V}{\partial{\mathrm N}_2\partial\nu^3}
     = {\rm sign}(\nu_{r})\frac{\partial^2 V}{\partial\nu_2\partial\nu_3};\\

     \frac{\partial^2 V}{\partial{\mathrm N}_1^2} = \frac{\partial^2 V}{\partial \nu_1^2};\\

     \frac{\partial^2 V}{\partial{\mathrm N}_1\partial{\mathrm N}_2}
     = \frac{\partial^2 V}{\partial\nu_1\partial\nu_2};\\

     \frac{\partial^2 V}{\partial{\mathrm N}_2^2} = \frac{\partial^2 V}{\partial \nu_2^2};\\

     {\mathrm d}^2V(\vec{e}_1, \vec{\nu}_0^\top)
     = {\rm sign}(\nu_{r})\left(\frac{\partial^2 V}{\partial x_1\partial\nu_1}\nu_{z}
     - \nu_{r}\frac{\partial^2 V}{\partial x^1\partial\nu^3}\right);\\

     {\mathrm d}^2V(\vec{e}_3, \vec{\nu}_0^\top)
     = {\rm sign}(\nu_{r})\left(\frac{\partial^2 V}{\partial x_3\partial\nu_1}\nu_{z}
     - \nu_{r}\frac{\partial^2 V}{\partial x^3\partial\nu^3}\right);\\
  \end{cases}
\label{N1}
\eqno(2.12.3.1)\]
\[
  \begin{cases}
     \frac{\partial^2 V}{\partial x_1\partial\nu_1} = -\mu B_{r,r};\\

     \frac{\partial^2 V}{\partial x_1\partial\nu_2} = 0;\\

     \frac{\partial^2 V}{\partial x_3\partial\nu_1} = -\mu B_{r,z};\\

     \frac{\partial^2 V}{\partial x_3\partial\nu_2} = 0;\\

     \frac{\partial^2 V}{\partial x_1\partial\nu_1}\nu_z
     - \nu_{r}\frac{\partial^2 V}{\partial x^1\partial\nu^3}
     = -\mu (B_{r,r}\nu_z - B_{r,z}\nu_{r});\\

     \frac{\partial^2 V}{\partial x_3\partial\nu_1}\nu_z
     - \nu_{r}\frac{\partial^2 V}{\partial x^3\partial\nu^3}
     = -\mu (B_{r,z}\nu_z - B_{z,z}\nu_{r});\\
  \end{cases}
\label{N2}
\eqno(2.12.3.2)\]
\[ \begin{cases}
      \frac{\partial^2 V}{\partial x_1^2}
      = -\mu\nu^z B_{z,rr} + \mu\nu^r B_{z,zr} - \mu\nu^r\frac1{r}\left(B_{z,z} + \frac2{r}B_r\right);\\
      \frac{\partial^2 V}{\partial x_1\partial x_3}
      = -\mu\nu^z B_{z,rz} - \mu\nu^r B_{z,rr};\\
      \frac{\partial^2 V}{\partial x_3^2}
      = -\mu\nu^z B_{z,zz} - \mu\nu^r B_{z,rz};\\
   \end{cases}
\label{N3}
\eqno(2.12.3.3)\]
\[ \begin{cases}
      \mu (B_{r,z}\nu_{r} - B_{r,r}\nu_z) = M g + \frac{\mu B_r}{r}\nu_z;\\
      \mu (B_{z,z}\nu_{r} - B_{r,z}\nu_z)  = M\omega^2 r - \frac{\mu B_r}{r}\nu_{r};\\
   \end{cases}
\label{N4}
\eqno(2.12.3.4)\]

Then the stability conditions of the relative equilibrium in 2.11 can be reduced to the form
\[ \begin{cases}
      \lambda = \mu\frac{B_z}{\nu_z} + \frac{\omega C_2}{\nu_z} - I_\perp\omega^2 > 0;\\
     A > 0,\quad C > 0,\quad A C - B^2 > 0;\\
   \end{cases}
\label{N5}
\eqno(2.12.3.5)\]
\[  A =
    M\omega^2\frac{3 M r_0^2 - I_\perp\nu_r^2}{I_\perp\nu_r^2 + M r_0^2}
    - \mu\nu^z B_{z,rr} + \mu\nu^r B_{z,zr} - \mu\nu^r\frac1{r}\left(B_{z,z} + \frac2{r}B_r\right)
\label{N6a}
\eqno(2.12.3.6a)\]
\[ -\frac{\left(2\frac{I_\perp\nu_r p_0}{I_\perp\nu_r^2 + M r_0^2}
      \left(\nu_{z}\omega + \lambda_2\right)
   + \left(M g + \frac{\mu B_r}{r}\nu_z\right)\right)^2}
   {\lambda  + \frac{I_\perp^2\nu_r^2}{I_\perp\nu_r^2 + M r_0^2}
     \left(\nu_{z}\omega + \lambda_2\right)^2
   + \nu_r^2 I_\perp\omega^2}
\]
\[  B =  -\mu\nu^z B_{z,rz} - \mu\nu^r B_{z,rr}
\label{N6b}
\eqno(2.12.3.6b)\]
\[ - \frac{\left(2\frac{I_\perp\nu_r p_0}{I_\perp\nu_r^2 + M r_0^2}\left(\nu_{z}\omega + \lambda_2\right)
   + \left(M g + \frac{\mu B_r}{r}\nu_z\right)\right)}
   {\lambda + \frac{I_\perp^2\nu_r^2}{I_\perp\nu_r^2 + M r_0^2}\left(\nu_{z}\omega + \lambda_2\right)^2
   + \nu_r^2 I_\perp\omega^2}\left(M\omega^2 r - \frac{\mu B_r}{r}\nu_{r}\right)
\]
\[ C = -\mu\nu^z B_{z,zz} - \mu\nu^r B_{z,rz}
   - \frac{\left(M\omega^2 r - \frac{\mu B_r}{r}\nu_{r}\right)^2}
     {\lambda + \frac{I_\perp^2\nu_r^2}{I_\perp\nu_r^2 + M r_0^2}
     \left(\nu_{0z}\omega + \lambda_2\right)^2
   + \nu_r^2 I_\perp\omega^2}
\label{N6c}
\eqno(2.12.3.6c)\]
where values of the variables $\nu_r,\nu_z,\omega$
can be found from (2.12.2.3) and $\lambda_2,\lambda,\vec{\pi}_0$
from (2.12.3).

\section{ The Orbitron }
\label{Orbitron}

The most simple example is the Orbitron.

As the field of the Orbitron (we mean generalized Orbitron see in \cite{GOZ})
we assume the field axially-symmetric about an axis $z$
and mirror symmetric respect to the plane vertical to axis $z$.

Let's show that there are stable relative equilibrium
that are located in the plane of $z$ (equatorial plane).

In this case without gravity force
\[ V(\vec{x},\vec{\nu}) = -\mu B_i(\vec{x})\nu^i
  = -\mu\langle\vec{B}(\vec{x}),\vec{\nu}\rangle
\label{O1}
\eqno(2.13.1)\]

In view of the mirror symmetry the supporting point is given by
\[
   \begin{cases}
      \vec{x}_0 = r_0\vec{e}_1;\\
      \vec{p}_0 = M\omega r_0\vec{e}_2;\\
      \vec{\nu}_0 = \sigma\vec{e}_3,\quad \sigma=\pm 1; \\
      \vec{\pi}_0 = \sigma C_2\vec{e}_3 = \pi_3\vec{e}_3 = \pi_0\vec{e}_3;\\
   \end{cases}
\label{O2}
\eqno(2.13.2)\]
and, therefore,
\[
   \begin{cases}
      \lambda_2 = \sigma\left(\omega - \frac1{I_\perp}\pi_0\right);\\
      \lambda = \sigma\mu B_z + \omega(\pi_0 - I_\perp\omega)
   \end{cases}
\label{O3}
\eqno(2.13.3)\]
\[
   \begin{cases}
      B_{z,z} = 0;\\
      B_{z,r} = -\sigma\frac{M}{\mu}\omega^2 r_0;\\
      \vec{B} = B_z\vec{e}_3;\\
   \end{cases}
\label{O4}
\eqno(2.13.4)\]

Now we consider the sufficient conditions.

By using all previous simplifications of (2.12.3.5, 2.12.3.6a-6c) we obtain
\[
   \begin{cases}
      \lambda = \sigma\mu B_z + \omega\pi_0 - I_\perp\omega^2 > 0;\\
      B = 0;\\
   \end{cases}
\label{O5}
\eqno(2.13.5)\]

\[ A = 3M\omega^2 - \mu\sigma B_{z,rr}  > 0
\label{O6a}
\eqno(2.13.6a)\]
\[ C = -\sigma\mu B_{z,zz} - \frac{\left(M\omega^2 r\right)^2}{\lambda} > 0
\label{O6c}
\eqno(2.13.6c)\]

From (2.12.3.6c) follows $-\sigma B_{z,zz} > 0$
then the conditions (2.12.3.6a,2.12.3.6c) can be written in the form
\[
   \begin{cases}
      -\sigma B_{z,zz} > 0;\\
      -\sigma\left(3\frac1r B_{z,r} + B_{z,rr}\right)
      = -\sigma\left( - B_{z,zz} + 2\frac1r B_{z,r}\right) > 0;\\
      \omega\pi_0 > -\sigma\mu B_z + I_\perp\omega^2 + \frac{\mu (B_{z,r})^2}{-\sigma B_{z,zz}};\\
   \end{cases}
\label{O7}
\eqno(2.13.7)\]

These conditions are equivalent to the conclusions of the article \cite{GOZ}.

\subsection{ Dipoletron }

In standard Orbitron \cite{ZubOrb13} as the sources of the magnetic field
the magnetic poles were used. The results shown above demonstrates
that other kinds of the field sources are also allow the stable confinement.

A special interest are the sources of the field in the form of magnetic dipoles,
because of recognized model of a field.
Furthermore, it is known that stable dynamic configurations not realizable
in the system of magnetic dipoles that interact by magnetic forces only
the (so-called "problem $\frac1{R^3}$ \cite{GinzbNucForce47}).
This circumstance adds the interest to such choice of the model field.

Let's consider a system with two magnetic field sources
that are the magnetic dipoles located on the axis $z$ at points $z=\pm h$
and with equal magnetic moments oriented along the same axis $z$.
Obviously, the field of this system is axially symmetric about axis $z$
and mirror-symmetric with respect to the plane $z=0$.

{\bf Remark 6}. In our system, we have outside forces that keep the dipoles-sources
in a predetermined position.

The field of the magnetic dipole is well known
\[ \vec{B} = \frac{\mu_0}{4 \pi}\left(
          \frac{ 3\langle\vec{m},\vec{R}\rangle \vec{R}}{R^5}
         - \frac{\vec{m}}{R^3}\right),
\label{P1}
\eqno(2.13.1.1)\]
where $\vec{m}$ vector of the magnetic moment,
and $\vec{R}$ is the radius-vector from the dipole to the point of field observation.

For the dipoles located on the axis $z$ in the points $\pm h$ in the components of Cartesian system we have
\[
  \begin{cases}
     B^{\pm}_{C} = 3q\frac{(x_3\mp h) x_C}{D_{\mp}^{5/2}}, \quad C = 1,2;\\
     B^{\pm}_{3} =  q\frac{2(x_3\mp h)^2 - (x_1^2 + x_2^2)}{D_{\mp}^{5/2}},
     \quad D_{\pm} = x_1^2+x_2^2+(x_3\mp h)^2;
  \end{cases}
\label{P2}
\eqno(2.13.1.2)\]
where $q = \frac{\mu_0}{4 \pi}|\vec{m}|$ --- "magnetic charge" equivalent to a pole of a magnetic dipole.

By using (2.13.1.2) and take a derivative of $B_z$ with respect to $r$ at $z=0$
we can express all quantities of interest.
\[
  \begin{cases}
     B_{z} =  2q (2h^2 - r_0^2){D_{0}^{-5/2}}; \\
     B_{z,r} = B_{r,z} = r\beta_{,z} = -6q r_0(-r_0^2 + 4 h^2)D_0^{-7/2}; \\
     B_{z,zz} = 6q(3 r_0^4 - 24 r_0^2 h^2 + 8h^4)D_0^{-9/2}; \\
     \frac{3}{r}\frac{\partial B_z}{\partial r}
     + \frac{\partial^2 B_z}{\partial r^2}
     = -6q(r_0^4 - 28 r_0^2 h^2 + 16 h^4)D_0^{-9/2},
  \end{cases}
\label{P3}
\eqno(2.13.1.3)\]
where $D_0=r_0^2+h^2$.

Then from the first two conditions (2.13.7) we obtain the geometric conditions for the Dipoletron system
\[
  \begin{cases}
     3(\frac{r_0}{h})^4 - 24 (\frac{r_0}{h})^2 + 8 < 0; \\
      (\frac{r_0}{h})^4 - 28 (\frac{r_0}{h})^2 + 16 > 0;
  \end{cases}
\label{P4}
\eqno(2.13.1.4)\]
or
\[ 2\sqrt{1 - \sqrt{5/6}} < \frac{r_0}{h} < \sqrt{9 - \sqrt{65}}
\label{P5}
\eqno(2.13.1.5)\]

with fulfilling of geometric conditions (2.13.1.5) the third condition (2.13.7) can alway be satisfied.

\section{ Levitation of the Orbitron }
\label{Levitation}

Let's explore the possibility of the levitation of a dipole in an axially symmetric magnetic field.

As it was shown above, the field of the Orbitron can provide
the equilibration of the centrifugal force with stability.
However, the studies show that this field is ill-suited for the equilibration of gravity force.
Therefore, add the field that linearly depends of coordinates in the system.

Let
\[ \vec{B}(\vec{x}) = \vec{B}^L(\vec{x}) + \vec{B}^O(\vec{x})
\label{Q1}
\eqno(2.14.1)\]
where $\vec{B}^L$ --- the magnetic field is linearly dependent on the coordinates,
and $\vec{B}^O$ --- the field that mirror-symmetric with respect to the plane $z=0$
(of course, both fields axially-symmetric).

It is expected that the $\vec{B}^L$ field for the most part is intended
for the compensation of the gravity force,
and $\vec{B}^O$ field for the compensation of the centrifugal force.

Let's consider the relative equilibria that are spatially observed in $z=0$.

The main properties of these fields are
\[
  \begin{cases}
     B^O_r|_{z=0} = 0\longrightarrow (B^O_{r,r} = 0)\& (B^O_{r,rr} = 0)|_{z=0}\\
     \longrightarrow (B^O_{z,z} = 0)\& (B^O_{z,rz} = 0)|_{z=0}; \\
    {\mathrm d}^2 \vec{B}^L = 0;\\
  \end{cases}
\label{Q2}
\eqno(2.14.2)\]

We have (see \cite{GeimDiamagLev01})
\[
  \begin{cases}
     B^L_z = B_0 + B'z;\\
     B^L_r = -\frac12 B'r;
  \end{cases}
\label{Q3}
\eqno(2.14.3)\]
\[
  \begin{cases}
     B^L_{z,z} = B';\\
     B^L_{r,z} = B^L_{z,r} = 0;\\
     B^L_{r,r} = -\frac12 B';
   \end{cases}
\label{Q4}
\eqno(2.14.4)\]

Then
\[
  \begin{cases}
     B_{z,z} = B';\\
     B_{r,z} = B^O_{r,z};\\
     B_{r,r} = -\frac12 B';
  \end{cases}
\label{Q5}
\eqno(2.14.5)\]

\subsection{ Necessary conditions of the relative equilibrium }

Equation of equality of the forces can be written as
\[
  \begin{cases}
     \nu^z  + \beta\nu^r  = \kappa;\\
     \beta\nu^z  - \frac12 \nu^r  = -\kappa\xi^2;\\
     \nu_r^2 + \nu_z^2 = 1
  \end{cases}
\label{R1}
\eqno(2.14.1.1)\]
where
\[
  \begin{cases}
     \beta = \frac{B^O_{r,z}}{B'};\\
     \kappa = \frac{M g}{\mu B'};\\
     \xi^2 = \frac{\omega^2 r}{g};\\
  \end{cases}
\label{R2}
\eqno(2.14.1.2)\]

From here
\[ \nu_r = \frac{1}{1 + \beta^2}\left(\beta\kappa
        \pm \sqrt{1 + \beta^2 - \kappa^2}\right)
\label{R3}
\eqno(2.14.1.3)\]
\[  \xi^2 = \frac1{\kappa}\left(\frac12 + \beta^2\right)\nu_r - \beta
\label{R4}
\eqno(2.14.1.4)\]

From the equation
\[ \nu_r = \frac{\mu}{\lambda}B_r
\label{R5}
\eqno(2.14.1.5)\]
and (2.14.3) we have
\[ \nu_r = -\frac{M g r}{2\kappa\lambda}\longrightarrow
   {\rm sign}(\nu_r) = -{\rm sign}(\kappa)
\label{R6}
\eqno(2.14.1.6)\]

Thus the sign of $\nu_r$ should be opposite to the sign of $\kappa$.

Therefore, in the expression (2.14.1.4) 1-st member
and must be negative, then the sign of the $\beta$
must be the negative.
\[  \beta < 0
\label{R7}
\eqno(2.14.1.7)\]

Furthermore, since vector $\vec{\nu}$ is completely determined
of the system (2.14.1.1), then the value of $\lambda$ is also defined by (2.14.1.6).

Therefore, the equation
\[ \lambda = \mu\frac{B_z}{\nu_z} + \frac{\omega C_2}{\nu_z} - I_\perp\omega^2 > 0;\\
\label{R8}
\eqno(2.14.1.8)\]
should not be regarded as an equation for $\lambda$,
but rather as a limitation for $B_z$ and $\pi$.

From (2.14.1.3) and (2.14.1.4) we have
\[  \xi^2 = - \frac{\beta}{2(1 + \beta^2)}
          \pm \frac{\frac12 + \beta^2}{1 + \beta^2}\frac{\sqrt{1 + \beta^2 - \kappa^2}}{\kappa}
\label{R4a}
\eqno(2.14.1.4a)\]

If $|\kappa|=1$ then this value with minus sign before the square root becomes negative.

Obviously (because of $\beta < 0$) the choice of the plus sign is preferred
when $\kappa>0$, and, conversely, for $\kappa<0$ it is necessary to choose the minus sign.

Otherwise, in the most interesting region $\kappa\approx 1$ we receive a negative value for the $\xi^2$.

Thus
\[  \xi^2 = - \frac{\beta}{2(1 + \beta^2)}
          + \frac{\frac12 + \beta^2}{1 + \beta^2}\frac{\sqrt{1 + \beta^2 - \kappa^2}}{|\kappa|}
\label{R4b}
\eqno(2.14.1.4b)\]

Then
\[ \nu_r = \frac{\kappa }{1 + \beta^2}\left(\beta
        + \frac{\sqrt{1 + \beta^2 - \kappa^2}}{|\kappa|}\right)
\label{R3a}
\eqno(2.14.1.3a)\]

Obviously, that a small excess of the $|\kappa|$ over one means that $\nu_r$ value will be small.
\[ |\kappa| = 1+\epsilon \longrightarrow \nu_r = O(\epsilon)(\epsilon>0)
\label{R9}
\eqno(2.14.1.9)\]
\[ \lambda = -\frac{M g r}{2\kappa\nu_r}\longrightarrow
   \frac{M g r}{\lambda} = O(\epsilon) > 0
\label{R6a}
\eqno(2.14.1.6a)\]

We also have
\[  \xi^2 = |\beta| + O(\epsilon)
\label{R10}
\eqno(2.14.1.10)\]

\subsection{ Sufficient conditions of stability for levitation of the Orbitron }

The expressions of $A,B,C$ take the form
\[
  \begin{cases}
      A = M\omega^2\frac{3 M r_0^2 - I_\perp\nu_r^2}{I_\perp\nu_r^2 + M r_0^2} - \mu\nu^z B_{z,rr}
   -\frac{\left(2\frac{I_\perp\nu_r p_0}{I_\perp\nu_r^2 + M r_0^2}
      \left(\nu_{z}\omega + \lambda_2\right)
   + M g \left(1 - \frac{\nu_z}{2\kappa}\right)\right)^2}
   {\lambda  + \frac{I_\perp^2\nu_r^2}{I_\perp\nu_r^2 + M r_0^2}
     \left(\nu_{z}\omega + \lambda_2\right)^2
   + \nu_r^2 I_\perp\omega^2};\\
      B =  - \mu\nu^r B_{z,rr}
   - \frac{\left(2\frac{I_\perp\nu_r p_0}{I_\perp\nu_r^2 + M r_0^2}\left(\nu_{z}\omega + \lambda_2\right)
   + M g \left(1 - \frac{\nu_z}{2\kappa}\right)\right)}
   {\lambda + \frac{I_\perp^2\nu_r^2}{I_\perp\nu_r^2 + M r_0^2}\left(\nu_{z}\omega + \lambda_2\right)^2
   + \nu_r^2 I_\perp\omega^2}M g \left(\xi^2 - \frac{M g r}{4\kappa^2\lambda}\right);\\
      C = -\mu\nu^z B_{z,zz}
   - \frac{(M g)^2 \left(\xi^2 - \frac{M g r}{4\kappa^2\lambda}\right)^2}
     {\lambda + \frac{I_\perp^2\nu_r^2}{I_\perp\nu_r^2 + M r_0^2}
     \left(\nu_{z}\omega + \lambda_2\right)^2
   + \nu_r^2 I_\perp\omega^2};\\
  \end{cases}
\label{S1}
\eqno(2.14.2.1)\]

Given (2.14.1.6a,2.14.1.9,2.14.1.10), we get sufficient conditions in the form
\[
  \begin{cases}
      a = \frac{r}{M g} A
      = -\sigma\frac{\mu r}{M g}\left(\frac3r B_{z,r} + B_{z,rr}\right) - O(\epsilon)>0;\\
      b = \frac{r}{M g} B = O(\epsilon);\\
      c = \frac{r}{M g} C =  - \sigma\frac{\mu r}{M g}B_{z,zz} - O(\epsilon)>0;\\
      a c - b^2 > 0;\\
      \lambda = \sigma\mu B_z + \omega\pi_0 - I_\perp\omega^2 \gg \frac{M g r}{2};\\
  \end{cases}
\label{S2}
\eqno(2.14.2.2)\]
at that $O(\epsilon)>0$.

From (2.14.2.2) and $O(\epsilon)>0$ it follows that geometric conditions are fulfilled
\[
  \begin{cases}
     -\sigma B_{z,zz} > 0;\\
     -\sigma\left(\frac3r B_{z,r} + B_{z,rr}\right)
     = -\sigma\left( - B_{z,zz} + \frac2r B_{z,r}\right) > 0;\\
  \end{cases}
\label{S3}
\eqno(2.14.2.3)\]
that ensure compliance with sufficient stability conditions (2.14.2.2) for sufficiently small $\epsilon>0$.

Thus condition (2.14.2.3), together with dynamic condition
\[ \omega\pi_0 \gg -\sigma\mu B_z +  I_\perp\omega^2 + M g r
\label{S4}
\eqno(2.14.2.4)\]
gives a full set of conditions of the system stability,
i.e. Orbitron {\it can levitates}.

{\bf Remark 7}. Parameter $B_z$ to a certain extent is adjustable,
as contains the contribution of the field $\vec{B}^L$ that does not affect on the equilibrium
of the magnetic force and gravity force, and this parameter occures only in the (2.14.2.4).







\end{document}